\documentclass[fleqn,usenatbib]{mnras}

\usepackage{newtxtext,newtxmath}
\usepackage{graphicx}
\usepackage{xcolor}
\usepackage{color}
\usepackage{amsmath}
\usepackage{hyperref}
\usepackage{multirow}
\usepackage{notes2bib}
\usepackage{stfloats}
\usepackage{float}
\restylefloat{table}
\usepackage{caption} 
\usepackage{subcaption}

\usepackage{rotating}
\usepackage{pdflscape}
\usepackage{array}
\usepackage{hhline,booktabs,amsmath}
\usepackage{makecell}
\usepackage{placeins}
\usepackage{float}
\usepackage{comment}
\hypersetup{colorlinks=true,linkcolor=[rgb]{1.,0.2,0.2},citecolor=[rgb]{0.1,0.4,1.},filecolor=[rgb]{0.7,0.2,0.2},urlcolor=[rgb]{0.0,0.2,1.}}

\usepackage[T1]{fontenc}

\DeclareRobustCommand{\VAN}[3]{#2}
\let\VANthebibliography\thebibliography
\def\thebibliography{\DeclareRobustCommand{\VAN}[3]{##3}\VANthebibliography}

\usepackage{graphicx}	
\usepackage{amsmath}	

\def\INSPIRE{\mbox{{\tt INSPIRE}}}
\def\EINSPIRE{\mbox{{\tt E-INSPIRE}}}
\definecolor{darkgreen}{rgb}{0.05, 0.5, 0.3}
\definecolor{amber}{rgb}{1.0, 0.49, 0.0}
\newcommand{\chiara}[1]{{\textcolor{black}{#1}}}
\newcommand{\maxz}[1]{{\textcolor{black}{#1}}}

\title[Galactic Morpholography]{Galactic Morpholography: Linking present-day galaxy compactness to assembly history and relic formation in the COLIBRE simulations}
\author[Zemsch M. et al.]{
Maximilian Zemsch,$^{1}$\thanks{E-mail: maximilian.zemsch@physics.ox.ac.uk}
Claudia D.P. Lagos,$^{2,3}$
Chiara Spiniello,$^{4,1,5}$
Michele Cappellari,$^{1}$
\and \'Angel Chandro-G\'omez,$^{2}$
Katy L. Proctor,$^{6}$
Evgenii Chaikin,$^{7,8}$
Johanna Hartke,$^{9,10,11}$
Filip Hu\v{s}ko,$^{8}$
\and Sylvia Ploeckinger,$^{12}$
Alexander J. Richings,$^{13,14}$
Mahsa Sanati,$^{15}$
Joop Schaye$^{8}$
and James W. Trayford$^{16}$
\\
$^{1}$Sub-Dep. of Astrophysics, Dep. of Physics, University of Oxford, Denys Wilkinson Building, Keble Road, Oxford OX1 3RH, United Kingdom\\
$^{2}$International Centre for Radio Astronomy Research (ICRAR), M468, University of Western Australia, 35 Stirling Hwy, Crawley, WA 6009, Australia\\
$^{3}$Cosmic Dawn Center (DAWN), Denmark\\
$^{4}$European Southern Observatory,  Karl-Schwarzschild-Stra\ss{}e 2, 85748, Garching, Germany\\
$^{5}$INAF -  Osservatorio Astronomico di Capodimonte, Via Moiariello  16, 80131, Naples, Italy\\
$^{6}$The Oskar Klein Centre, Department of Physics, Stockholm University, AlbaNova University Center, 106 91 Stockholm, Sweden\\
$^{7}$Institute for Computational Cosmology, Department of Physics, University of Durham, South Road, Durham, DH1 3LE, UK\\
$^{8}$Leiden Observatory, Leiden University, PO Box 9513, 2300 RA Leiden, the Netherlands\\
$^{9}$Finnish Centre for Astronomy with ESO, (FINCA), University of Turku, 20014 Turku, Finland\\
$^{10}$Tuorla Observatory, Department of Physics and Astronomy, University of Turku, 20014 Turku, Finland\\
$^{11}$Turku Collegium for Science, Medicine and Technology (TCSMT), University of Turku, FI-20014 Turku, Finland\\
$^{12}$Department of Astrophysics, University of Vienna, Türkenschanzstrasse 17, A-1180 Vienna, Austria\\
$^{13}$Centre for Data Science, Artificial Intelligence and Modelling, University of Hull, Cottingham Road, Hull, HU6 7RX, UK\\
$^{14}$ E. A. Milne Centre for Astrophysics, University of Hull, Cottingham Road, Hull, HU6 7RX, UK\\
$^{15}$Department of Astronomy, University of Geneva, Chemin Pegasi 51, 1290 Versoix, Switzerland\\
$^{16}$Centre for Astrophysics Research, Department of Physics, Astronomy and Mathematics, University of Hertfordshire, College Lane, Hatfield, AL10 9AB, UK\\
}
\date{Accepted XXX. Received YYY; in original form ZZZ}

\pubyear{\the\year{}}

\begin{document}
\label{firstpage}
\pagerange{\pageref{firstpage}--\pageref{lastpage}}
\maketitle

\begin{abstract}
\noindent Compactness is widely used to identify candidate relic galaxies, the largely untouched survivors of the first compact massive galaxies. Yet relicness is fundamentally an evolutionary rather than structural property. Using the COLIBRE cosmological hydrodynamical simulations, we test whether present-day structure traces galaxy formation and assembly histories. Six complementary tracers (stellar age, metallicity, [Mg/Fe], sSFR, velocity dispersion and ex-situ mass fraction) vary coherently across the stellar mass--size plane and independently identify a characteristic compactness scale of $\log_{10}\!\left(\Sigma_{1.5,\mathrm{crit}}/(\mathrm{M}_\odot\,\mathrm{kpc}^{-1.5})\right)=9.75\pm0.10$. We term this mapping between present-day structure and assembly history {\it morpholography}. Combining this threshold with the observationally motivated Degree of Relicness (\({\rm DoR}>0.6\)), which quantifies early and rapid star formation, we separate compact from ancient galaxies and identify 267 relic candidates at their intersection. Ancient galaxies span nearly the full ex-situ-fraction range, from predominantly in-situ formation to substantial merger-driven growth. Relic candidates are concentrated at stellar masses of $\approx 10^{10}$--$10^{11}\,{\rm M_\odot}$, host black holes that lie $\approx 0.25$--$2.5$ dex above the median black-hole--stellar-mass relation, and are predominantly satellites ($76\%$), preferentially in cluster environments. Only $\approx 3\%$ show evidence of strong tidal stripping, suggesting that environment primarily preserves rather than creates relic-like systems. Approximately one third ($87/267$) are direct descendants of the $z=2$ massive quiescent galaxies identified in COLIBRE. Present-day structure thus retains a statistical fossil record of assembly history, supporting morpholography as a link between nearby galaxies and the early Universe.
\end{abstract}




\section{Introduction}
Massive galaxies are commonly thought to form through a two-phase evolutionary scenario: an early phase of rapid, in-situ star formation at $z \gtrsim 2$, followed by a prolonged period of mass and size growth driven by ex-situ accretion \citep{Naab+09, Oser+10}. This scenario is supported by high-redshift observations that have revealed a population of compact, massive “red nugget” galaxies \citep{Damjanov+09}, which are widely interpreted as the end-product of the first phase and the progenitors of present-day extended ellipticals \citep{Pulsoni21, Barbosa21}. While most of these systems undergo substantial structural evolution and size growth through mergers during the second phase, a small fraction is expected to largely avoid this stage and survive to the present day as relic galaxies \citep{Trujillo14}.  
\chiara{Observationally, relic galaxies are defined as compact, massive
systems whose inferred star-formation histories indicate that they
formed the bulk of their stellar mass rapidly by $z\simeq2$ and
experienced little subsequent mass or size growth
\citep{Ferre-Mateu+17,Spiniello+24,Mills25}. 
The choice of $z=2$ is not a fundamental boundary, but approximately
marks the epoch by which compact massive galaxies were already in
place and after which ex-situ growth is expected to become increasingly
important \citep{Naab+09,Oser+10,Zolotov15}.} 
Identifying and characterising these relic systems, whose stellar populations are nearly as old as the Universe, provides a unique opportunity to constrain the earliest phases of galaxy assembly with a level of detail only accessible in the nearby Universe \citep{Spiniello20_Pilot} and to investigate the co-evolution between host galaxy and central black hole (BH) \citep{Ferre-Mateu+15, Walsh+15, Yildirim+15}.

\chiara{Old stellar populations alone, however, do not uniquely establish
relicness. Present-day giant ellipticals are dominated by old,
metal-rich, and $\alpha$-enhanced stellar populations
\citep[e.g.][]{Thomas+05,McDermid15}, yet simulations indicate that
they have undergone substantial late-time ex-situ growth
\citep{Rodriguez-Gomez+16,Segers+2016,Clauwens+2018,Pulsoni21}.
Their central stellar populations may therefore preserve the signature
of rapid early formation \citep{Sarzi+18,Barbosa21}, while the galaxy
as a whole remains the product of significant merger-driven assembly
\citep{Qu2017}. Consequently, old ages and enhanced
$\alpha$-element abundances, particularly when measured in central
regions, do not necessarily imply that a galaxy avoided subsequent
accretion. }
This distinction between early stellar formation and subsequent structural assembly is central to separating genuinely untouched relics from old, massive merger remnants.

\chiara{The most promising candidates for massive relic galaxies are passive
compact massive galaxies (CMGs; \citealt{Buitrago+18_compacts,
Tortora+18_UCMGs,Scognamiglio20}), which occupy the small-size,
high-density tail of the stellar mass--size plane \citep{Mills25}. Their precise definition varies between studies, which adopt different size-,
density-, or percentile-based thresholds. 
However, compactness is a structural property, whereas relicness is
inherently evolutionary, requiring an early formation epoch and
minimal subsequent merger-driven growth
\citep{Trujillo14,Ferre-Mateu+17,Spiniello+21}.}  
Distinguishing between these two concepts is therefore essential: not all passive CMGs are relics, although relic galaxies are expected to be compact, as they largely avoid the size-growth phase. Observational efforts have made significant progress in identifying CMGs and separating them into relic, those with an incredibly early and rapid star formation, and non-relic populations, showing a somewhat more time-extended star formation history. In particular, the \INSPIRE\ and \EINSPIRE\ projects \citep{Spiniello+24, Mills25} have assembled the largest samples of CMGs with detailed constraints on their stellar populations \citep{Spiniello+21, Spiniello+24}, also including the Initial Mass Function slope \citep{Martin-Navarro+23, Maksymowicz-Maciata24}, kinematics \citep{DAgo23}, and structural and photometric properties \citep{Tortora25,Spiniello25,Spiniello26}. 

Observational studies are necessarily limited to present-day observables and rely on selection criteria that are, at least in part, empirical. As a result, any possible connections between structural compactness and the underlying evolutionary history remain indirect. 
Furthermore, a quantitative definition of compactness remains ambiguous in observations \citep{Charbonnier+17_compact_galaxies}. Galaxies populate a continuous distribution in the mass–size plane, without a clear separation between compact and extended systems. As a result, compactness is often defined using simple size thresholds or percentile-based cuts, which are straightforward to apply but lack a clear physical motivation. Alternative parametrisations to quantify compactness, like the \maxz{velocity dispersion} $\sigma_{\rm e}$ of the stars, or equivalently the ratio $\sigma_{\rm e}^2\propto M/R_{\rm e}$ \citep{Cappellari+13_ATLAS3D_XX, Cappellari16}, or the surface-density within 1 kpc \citep[e.g.][]{Cheung2012, Barro13, Baldry21}, capture empirical trends but remain tied to observational calibrations rather than underlying evolutionary processes.

In this work, we take a complementary approach using cosmological hydrodynamical simulations to directly link galaxy structure to evolutionary history. We show that structural properties, and most notably compactness, encode information about star formation timescales and chemical enrichment, effectively acting as a structural fossil record of galaxy evolution. We refer to this mapping between morphology and evolutionary history as {\it morpholography}, in which the mass–size plane does not simply describe galaxy structure, but also encodes information about the timing and duration of galaxy formation. 

A growing number of studies have explored the formation and survival of compact massive galaxies and relic systems using cosmological simulations such as EAGLE and IllustrisTNG \citep[e.g.][]{Furlong+17, Wellons16, PeraltaDeArriba+16, Flores-Freitas22, Moura24}. These works have provided important insights into the physical mechanisms governing the emergence of compact systems and the conditions under which a subset of them can survive as relic galaxies. In particular, they highlight the central role of merger histories, stellar halo properties, and internal kinematics. 
However, despite these advances, simulation-based studies have been inherently limited by trade-offs between volume and resolution. High-resolution runs, such as TNG50 \citep{Nelson+19}, provide detailed information on internal galaxy structure and stellar populations, but probe relatively small cosmological volumes, making it difficult to sample rare objects such as massive relic galaxies. Conversely, larger-volume simulations often lack the resolution required to accurately capture the internal dynamics and assembly histories of compact systems. These limitations have so far prevented a comprehensive and statistically robust characterization of CMGs and relics across cosmic time. 
\maxz{The COLIBRE suite of cosmological hydrodynamical simulations \citep{Schaye+26,Chaikin2026a} overcomes many of these limitations by combining larger cosmological volumes with dark matter supersampling, which suppresses the spurious transfer of energy from dark matter particles to stars while maintaining sufficient resolution to resolve the internal structure of compact galaxies.} This makes COLIBRE especially well suited for studying rare systems such as compact massive galaxies and relics in a statistically meaningful way, while simultaneously resolving their internal structure and evolutionary pathways. 

This paper is structured as follows. In Section~\ref{sec:colibre}, we describe the simulation and define the key quantities used in our analysis in section~\ref{sec:sample}. In Section~\ref{sec:SCGs}, we examine how compactness correlates with stellar population properties across the mass–size plane and derive a physically motivated compactness threshold. In Section~\ref{sec:relics}, we introduce the concept of ancient galaxies and combine this with compactness. In Section~\ref{sec:compact-relic-relation}, we identify and characterise simulated relic candidates as the overlap of compact and ancient galaxies, and explore possible quenching pathways. We discuss  results and  their implications in Section~\ref{sec:discussion} and presents our conclusions in Section~\ref{sec:conclusions}.

\section{The COLIBRE cosmological simulations}
\label{sec:colibre}

In this study, we use the COLIBRE suite of cosmological hydrodynamical simulations. \chiara{Below, we provide a general overview on the main features and characteristics, while we refer to \citet{Schaye+26} and \citet{Chaikin2026a} for a more complete description}.

\subsection{Simulation overview}
COLIBRE combines large comoving cosmological volumes (50–400 cMpc on a side, where cMpc denotes comoving megaparsecs) with relatively high mass resolution ($\approx 10^5 - 10^7 \, \rm{M}_\odot$ for both baryonic and dark matter), enabling, among other things, the simultaneous study of rare, massive systems and the internal structure of galaxies across a broad mass range. This combination is particularly important for the present work, which focuses on compact and intrinsically rare objects in the local Universe.

A key feature distinguishing COLIBRE from many contemporary large-volume cosmological hydrodynamical simulations (reviewed by e.g. \citealt{Vogelsberger2020, Crain+23}) is its multiphase treatment of the interstellar medium (ISM), which allows gas to cool efficiently to low temperatures, $\approx 10$~K,  without imposing a restrictive temperature or pressure floor that would artificially limit gas compression \citep{Ploeckinger+25}. This enables the formation of dense, cold gas phases and a realistic multiphase ISM, which leads to approximately the correct partition of atomic and molecular gas at global galaxy scales in the local Universe \citep{Schaye+26}, while also reproducing the relationship between star formation, gas metallicity, stellar surface density and the molecular and atomic gas content on kiloparsec scales \citep{Lagos+25}. 
The multiphase ISM treatment is particularly important for the present work, as the formation of compact galaxies is expected to depend sensitively on the ability of gas to collapse to high central densities. 

COLIBRE also incorporates physically motivated models for pre-supernova stellar feedback and core-collapse supernova feedback \citep{Schaye+26}. The stellar feedback regulates the collapse of dense gas into stars while suppressing numerical overcooling through stochastic thermal energy injection \citep{Chaikin+23}. Together with the multiphase ISM treatment, this enables a realistic description of star formation in dense environments, which is particularly relevant for the formation of compact galaxies.

Another important feature is the supersampling of dark matter. Since dark matter contributes much more mass than baryons, using the same number of dark matter and baryonic particles would make the individual dark matter particles substantially more massive, leading to spurious gravitational collisional heating of the stellar component through numerical two-body interactions. Unlike gas particles, stars cannot radiate this excess energy away, causing them to become dynamically hotter and artificially increasing galaxy sizes and scaleheights. COLIBRE therefore supersamples the dark matter by a factor of four, substantially suppressing this numerical artefact and thereby reducing the artificial expansion of galaxies, particularly relevant for compact and old systems \citep{Ludlow+23}.

In addition, COLIBRE includes an extended model for stellar evolution and chemical enrichment, tracking individual elements \citep{Correa+26} and dust grains \citep{Trayford+26}. The detailed chemical evolution modelling is essential for our analysis, given the observational evidence that relic galaxies exhibit enhanced [Mg/Fe] \citep{Spiniello+24, Mills25} \maxz{and silicon abundances \citep{Eftekhari+26}} tracing very short star formation timescales \citep{Thomas+05}.

{Black hole growth and AGN feedback are modelled self-consistently in COLIBRE. COLIBRE provides two AGN feedback models: the fiducial thermal model and a hybrid thermal+jet model \citep{Husko2026}. \citealt{Chandro-Gomez+26b} showed that the thermal model of AGN feedback in COLIBRE successfully reproduces the evolution of the number density of massive quenched galaxies at $z>2$, while the hybrid AGN feedback model leads to number densities that are too low compared to observations because supermassive black hole growth is effectively delayed relative to the thermal model. As such, we use the fiducial thermal AGN model throughout this work
 given the importance of the high-$z$ population of massive quenched galaxies in providing progenitors of $z=0$ relic galaxies.}

{The COLIBRE simulation suite has been calibrated to reproduce key observational constraints on galaxy populations, including the evolution of the galaxy stellar mass function and the star formation rate–stellar mass relation \citep{Chaikin+25}. Beyond these calibration targets, COLIBRE successfully reproduces the evolution of the galaxy size–mass relation \citep{Ludlow+26}, the mass–metallicity relation \citep{Sharda+26}, the observed morphological mix of galaxies in the local Universe \citep{ForouharMoreno+26}, and the evolution of the passive galaxy fraction and number density of massive quiescent galaxies over cosmic time \citep{Chandro-Gomez+26a,Chaikin+25}.}
These successes demonstrate that COLIBRE provides a realistic framework for studying galaxy formation and evolution across a wide range of physical processes and scales.

We focus on the COLIBRE run with a cosmological volume of $(200\,\mathrm{cMpc})^3$ and a dark matter and baryonic particle mass resolution of $\sim 10^6\,\rm{M}_\odot$, which we refer to as L200m6 hereafter. The comoving gravitational softening length ($\epsilon_{\rm com}$) and the maximum proper gravitational softening length ($\epsilon_{\rm prop}$) in this simulation are $\epsilon_{\rm com}=1.8$~comoving kpc (ckpc) and $\epsilon_{\rm prop}=0.7$~physical kpc (kpc from hereafter), respectively. 

The large volume of this run also ensures sufficient statistics at the high-mass end, where compact galaxies are intrinsically rare, whereas the m6 resolution is sufficient to robustly resolve the stellar structure of galaxies across the stellar mass range considered here. In particular, \citet{Ludlow+26} showed that the sizes of galaxies with stellar masses $\gtrsim 10^9\,\rm M_{\odot}$ in COLIBRE at $z=0$ are well resolved at the m6 resolution,  
providing confidence that galaxy sizes and therefore structural compactness for galaxies above this mass threshold are reliably modelled. The adopted resolution also allows robust measurements of stellar masses and stellar population properties.

\subsection{Halo catalogues and merger trees}

The COLIBRE simulations provide snapshot outputs at a sequence of redshifts spanning the evolution of the Universe from the initial conditions to $z=0$. Haloes and subhaloes are identified, and their merger trees are built using the history-based HBT-HERONS code \citep{ForouharMoreno+25}, an updated version of HBT+ \citep{Han+18}, which assigns each galaxy a unique TrackID that enables robust tracking of progenitors and descendants across successive snapshots. HBT-HERONS is particularly effective at identifying subhaloes near halo centres \citep{ForouharMoreno+25} and reconstructing their merger histories with minimal contamination from numerical artefacts, such as spurious mass exchanges and erroneous progenitor–descendant links \citep{Chandro-Gomez2025a}, compared to other commonly used algorithms. This robustness is achieved through the explicit use of temporal information and is especially important for the present work, which relies on merger histories and ex-situ stellar mass fractions to distinguish different galaxy assembly pathways. Galaxy properties are measured using the Spherical Overdensity and Aperture Processor (SOAP; \citealt{McGibbon_2025}), which provides halo and subhalo catalogues for each snapshot together with a range of derived galaxy properties for every simulation snapshot.

The stellar ex-situ mass fraction is defined as the fraction of stellar mass formed outside the main progenitor branch and subsequently accreted through mergers. In practice, stellar particles are classified as ex-situ if, in the snapshot immediately following their formation, they are bound to a galaxy with a different TrackID than their $z=0$ descendant. Unless stated otherwise, we use this definition based on the full merger history throughout this work. For comparison, we also consider an alternative definition that excludes mergers during the early formation phase ($z>2$). In this case, stellar mass formed in progenitor galaxies with a SnapshotIndexOfDeath < 76 in HBT-HERONS is no longer classified as ex-situ.

\section{Sample selection and derived quantities}
\label{sec:sample}
Our starting sample comprises 129,850 simulated galaxies (both central and satellites) at redshift $z=0$ with stellar masses $M_\star \ge 10^9\,{\rm \rm{M}_\odot}$, where $M_\star$ denotes \maxz{the total mass of bound stellar particles enclosed within a 3D exclusive spherical aperture of radius 50 kpc centred on the most bound particle}. This aperture is chosen to encompass the stellar component of massive galaxies while excluding diffuse or unbound material at large radii. 
Finally, galaxy sizes are characterised by the 3D stellar half-mass radius, defined as the radius enclosing half of the stellar mass within the adopted 50 kpc aperture. This differs from the projected half-light radius commonly used in observations but can be measured directly in simulations without assumptions about stellar population synthesis or dust attenuation.

For each galaxy, we retrieve mean stellar age, instantaneous star formation rate, stellar metallicity, magnesium and iron abundances, and the most massive black hole particle mass \maxz{within the chosen aperture from the SOAP catalogues.} Stellar metallicities are computed from the stellar metal mass fraction as $[\text{Z/H}] \equiv \log_{10}(\rm Z_\star/\rm Z_\odot)$,
adopting $\rm Z_\odot=0.0134$ \citep{Asplund2009}, while abundance ratios are expressed in the standard logarithmic notation relative to the corresponding solar values as \mbox{$[{\rm Mg/Fe}] = \log_{10}\left(N_{\rm Mg}/N_{\rm Fe}\right)_\star - \log_{10}\left(N_{\rm Mg}/N_{\rm Fe}\right)_\odot$},
adopting $\log_{10}[(N_{\rm Mg}/N_{\rm Fe})_\odot] = 0.10$ \citep{Asplund2009}. 
These quantities trace the timing, duration, and ``burstiness'' of the star-formation history. In particular, magnesium is produced almost exclusively by core-collapse supernovae, whereas a substantial fraction of iron is released on longer timescales by type Ia supernovae \citep{Tinsley1979, Matteucci1986, Kobayashi+2020}. Consequently, stellar [Mg/Fe] provides a sensitive probe of the duration of star formation, with enhanced values indicating rapid stellar formation and quenching \citep{Thomas+99, Thomas+05}. 

We also retrieve the specific star formation rate (sSFR; $\equiv {\rm SFR}/M_{\star}$, where SFR and $M_{\star}$ are the instantaneous star formation rate and stellar mass, respectively), indicating the level of ongoing star formation activity relative to the galaxy mass. All these quantities are crucial for the characterisation of passive compact galaxies and relics. Beyond that, we consider the stellar velocity dispersion ($\sigma_{\star}$), which provides a dynamical measure of the stellar structure of galaxies. We compute $\sigma_\star$ as the root mean square of the diagonal components of the cylindrical (vertical) stellar velocity dispersion tensor {(i.e. $r-\phi-z$)} within the aperture of one stellar half mass radius. Recent analyses of the COLIBRE simulations have shown that galaxy morphology correlates strongly with internal kinematics and stellar population properties, with more spheroidal systems exhibiting higher stellar velocity dispersions, older stellar populations, higher black hole masses, and larger ex-situ fractions \citep{ForouharMoreno+26}. Since relic galaxies are expected to represent the most extreme compact and dynamically evolved systems, $\sigma_{\star}$ provides an additional tracer of their structural and evolutionary state.
The \INSPIRE\ and \EINSPIRE\ surveys have indeed reported that relics have a larger integrated stellar velocity dispersion than non-relics and normal-sized galaxies of similar stellar masses \citep{Spiniello+24, Mills25}. In this context, however, it should be noted as pointed out by \citet{ForouharMoreno+26} that the one-dimensional velocity dispersion adopted here is not directly equivalent to the aperture- and projection-dependent line-of-sight velocity dispersions measured observationally, as it does not include effects such as different orientations and beam smearing. 

We adopt $r$-band luminosity-weighted quantities where available in the SOAP catalogue for properties that are directly inferred from galaxy light in observations, such as stellar age and velocity dispersion. For quantities without a luminosity-weighted counterpart in the catalogue, such as the chemical abundances, we use the corresponding stellar mass-weighted values rather than deriving luminosity-weighted estimates from stellar population synthesis models. We verified that using mass-weighted alternatives, where available, does not alter the qualitative trends presented below.

\maxz{We also assess the fraction of stellar mass formed ex-situ, as defined above, to quantify the relevance of mergers and interactions in the evolutionary past of the present-day simulated galaxies. Relic galaxies are expected to largely avoid such an accretion-growth channel, displaying low ex-situ mass fractions.}

In addition, we calculate the star formation history (SFH) of each galaxy in COLIBRE by binning all stellar particles in a galaxy at $z=0$ by their birth's lookback time, using time bins of $10$~Myr. Using the initial mass of stellar particles formed in a time bin and the bin size, we calculate the SFR \maxz{of the $z=0$ galaxy’s
progenitors} at different epochs.
These histories also allow us to reconstruct the cumulative stellar mass growth as a function of cosmic time and to derive characteristic formation times and mass fractions used in the relic analysis below.
As described above, we retrieve black hole masses from the SOAP catalogues for COLIBRE. However, the black hole masses of passive satellite galaxies are affected by a known numerical artefact. In COLIBRE, black holes are repositioned onto nearby gas particles \citep{Bahe+22}, rather than the nearest collisionless particle, because the latter procedure would be considerably more computationally expensive. \maxz{Consequently, in gas-poor galaxies, black holes can become displaced from their host galaxies and may no longer be identified as belonging to them. This issue primarily affects satellite galaxies, and to a lesser extent backsplash centrals that have previously experienced environmental stripping. In total, 13402 out of 129850 galaxies (10.3\%) in the sample are affected by this.}
\maxz{To mitigate this problem, we trace galaxies with a reported black hole mass of zero at $z=0$ backwards through successive simulation snapshots using their TrackIDs, and adopt the black hole mass from the last snapshot in which a non-zero mass is associated with the galaxy.} We then adopt this value for the corresponding $z=0$ descendant. This procedure provides a conservative lower limit on the black hole mass that the galaxy would have in the absence of the numerical artefact, because it neglects any subsequent growth through gas accretion or mergers with other satellite galaxies. Nevertheless, it allows these systems to be retained in the analysis and their black hole masses to be compared consistently with those of the overall galaxy population at $z=0$.

Throughout this work, we focus only on simulated galaxies at $z=0$ with a stellar mass $\ge 10^9\,\rm M_{\odot}$, unless otherwise stated.

\section{Simulated Compact Galaxies (SCGs)}
\label{sec:SCGs}

\begin{figure*}
\includegraphics[width=\textwidth]{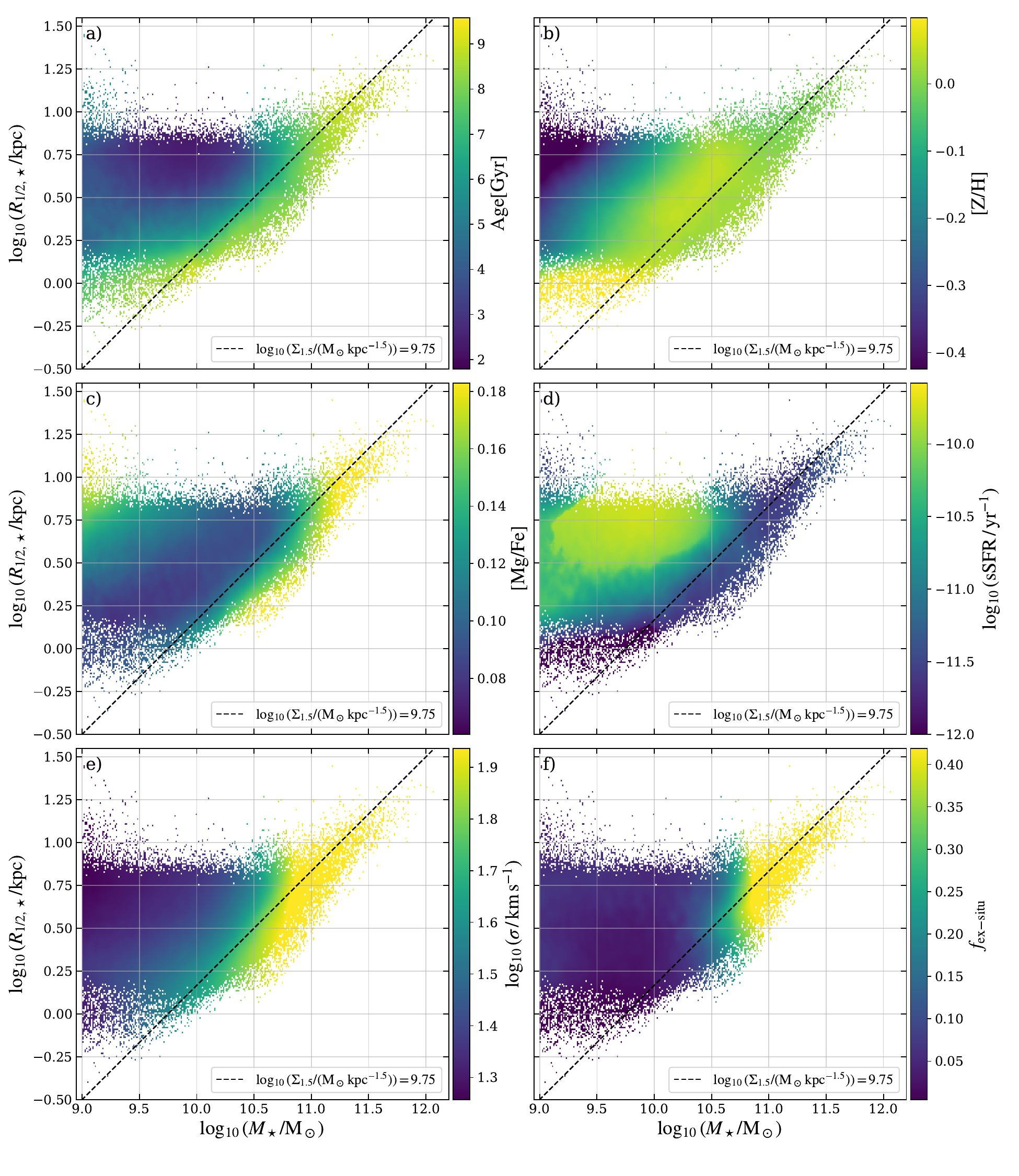}
\caption{Stellar mass–size plane at $z=0$ for all COLIBRE galaxies with stellar masses $M_\star \ge 10^9\,{\rm \rm{M}_\odot}$ coloured by different stellar-population diagnostics: (a) luminosity-weighted mean stellar age; (b) stellar metallicity; (c) stellar [Mg/Fe]; (d) specific star formation rate \maxz{(vanishing sSFR values leading to infinities in the logarithm are floored at $\log_{10} (\mathrm{sSFR}/\mathrm{yr}^{-1}) = -12$)}; (e) velocity dispersion; (f) ex-situ mass fraction. A LOESS regression is applied to highlight the main trends. The dashed line shown in all panels indicates a line of constant $\Sigma_{1.5}$, shown as a reference.}
\label{fig:fourpanels_masssize}
\end{figure*}

The central question of this work is \maxz{how} the present-day structure of galaxies encodes information about their evolutionary histories. We therefore investigate how the stellar population and assembly-history indicators introduced above are organised across the stellar mass–size plane. By combining directly observable properties with the ex-situ mass fraction, which can be measured explicitly from the simulated assembly histories, we can link observable trends to their underlying formation pathways and thereby provide a physically motivated framework for comparison with observational studies. 
If compactness correlates systematically with quantities tracing star formation, chemical enrichment, and merger-driven growth, it may serve as an indicator of past evolution rather than a purely morphological descriptor.

In the following, we show that compactness is indeed closely connected to multiple independent tracers of galaxy evolution. This connection forms the basis of the morpholography framework introduced in this work and allows us to derive a physically motivated compactness threshold marking a transition between galaxy populations with different formation channels. 

Figure~\ref{fig:fourpanels_masssize} shows the stellar mass--size plane for simulated COLIBRE galaxies that we colour-coded by luminosity-weighted stellar age (a), stellar metallicity (b), stellar [Mg/Fe] (c), sSFR (d)\footnote{Vanishing sSFR values leading to infinities in the logarithm are floored at $\log_{10} (\mathrm{sSFR}/\mathrm{yr}^{-1}) = -12$ to ensure all galaxies are displayed.}, $\sigma_{\star}$ (e), and ex-situ mass fraction (f). 
In each panel, a LOESS (Locally Estimated Scatterplot Smoothing) regression \citep{Cleveland1979} is applied to highlight the underlying large-scale trends in the distribution, while suppressing local fluctuations caused by the intrinsic scatter of individual galaxies, as implemented in the \textsc{loess} package\footnote{\url{https://pypi.org/project/loess/}} \citep{Cappellari+13_ATLAS3D_XX}. 
This facilitates the visual identification of systematic gradients and their relation to galaxy structure.
\chiara{Throughout this paper, we employ both scatter plots and LOESS visualisations. Scatter plots are used to inspect the distribution of individual galaxies and identify outliers, while LOESS visualisations emphasise the underlying population-level behaviour by smoothing over local fluctuations. Consequently, the colour scale in the LOESS maps reflects the smoothed quantity rather than the full dynamic range of the underlying data, and its extrema should not be interpreted as the minimum and maximum values present in the sample.}

A striking feature of these maps is the emergence of a coherent diagonal transition across the stellar mass--size plane. Galaxies on the left side of this transition are predominantly extended, younger, actively star-forming systems with relatively low [Mg/Fe] enhancement and lower stellar velocity dispersions. Moving diagonally across the plane towards more compact systems (bottom right), galaxies become progressively older, more quenched, dynamically hotter, and increasingly $\alpha$-enhanced. 
The alignment of these trends across multiple independent diagnostics strongly suggests a close connection between galaxy structure and evolutionary history. 

The strong correlation between stellar population properties and galaxy compactness is well established observationally. Compactness can be quantified in several related ways, including the central density slope, bulge-to-total mass ratio, central stellar density, or central velocity dispersion. 
\maxz{Using the ATLAS\textsuperscript{3D} survey, \citet{Cappellari+13_ATLAS3D_XX} showed that below a characteristic stellar mass of $M_{\star, \rm crit}\approx2\times10^{11}\,\rm{M}_\odot$, the distribution of ETG properties tends to be constant along lines of roughly constant \maxz{stellar velocity dispersion within one effective radius, $\sigma_{\rm e}$, where $R_{\rm e}$ is the effective radius}. Through the virial relation, these correspond approximately to lines with $R_{\rm e}\propto M_\star$, or even better to lines parallel to the Zone of Exclusion, $R_{\rm e}\propto M_\star^{0.75}$.}
This result has since been confirmed with increasingly large integral-field spectroscopic surveys, including SAMI \citep{Scott2017, Barone2018}, and MaNGA \citep{Li2018, Lu2023DynPop2}. Related studies have also demonstrated a strong connection between quenching and the stellar mass density within the central 1~kpc \citep{Cheung2012, Barro13, Baldry21}. Collectively, these works show that measures of galaxy compactness, and in particular central density, are more powerful predictors of quenching than total stellar mass \citep[see also the recent review in][]{Cappellari2026}, a trend that is clearly reproduced by the COLIBRE simulations.

To guide the eye, we overplot a diagonal line corresponding to a surface-density--like scaling. 
At this stage, this line is not intended to represent a formally defined threshold but rather to highlight the apparent separation between two broad evolutionary regimes. However, we anticipate that 
this visually identified transition corresponds closely to a quantitatively derived compactness threshold (see Sec.~\ref{sec:comp_crit}). Changes in the colour-coded quantities follow the diagonal direction.
This is particularly evident in the age, metallicity, and [Mg/Fe] 
panels, where the gradients are organised roughly parallel to lines of constant compactness. 
We caution the reader that the intrinsic velocity-dispersion measure adopted here is not directly comparable to observed line-of-sight velocity dispersions (see Sec.~\ref{sec:colibre}), and therefore does not necessarily follow the observational $\sigma^2\propto M_\star/R_e$ scaling.
While these transitions are continuous rather than sharply bimodal, the systematic variation of stellar population properties across the plane suggests that increasing compactness is associated with earlier and more rapid star formation histories. This motivates the possibility of a physically meaningful compactness threshold separating a population of simulated compact galaxies (SCGs) sharing key characteristics with relic systems.

In contrast, the ex-situ mass fraction exhibits a qualitatively different behaviour. Among all analysed quantities, the ex-situ fraction is unique in directly tracing merger-driven assembly history rather than stellar-population properties. The ex-situ fraction varies primarily along the stellar-mass direction, with more massive galaxies generally displaying larger ex-situ fractions, as expected from the cumulative accretion of externally formed stellar material during galaxy growth. Similar trends between stellar mass, accreted stellar components, and merger-driven assembly have been found both in simulations and observations \citep[e.g.,][]{Clauwens+2018, Spavone20, Pulsoni21}. 
Crucially, the apparent increase of ex-situ fraction towards the compact regime is driven mainly by the most massive galaxies. These systems combine high ex-situ fractions with old, quenched, and chemically evolved stellar populations, suggesting an evolutionary pathway distinct from that of relatively lower-mass compact galaxies, having a lower ex-situ fraction. 
This indicates that compact galaxies do not constitute a homogeneous population, but instead comprise systems with different assembly histories.

\subsection{A physically motivated compactness criterion}
\label{sec:comp_crit}

\begin{figure*}
\centering

\includegraphics[width=\textwidth]{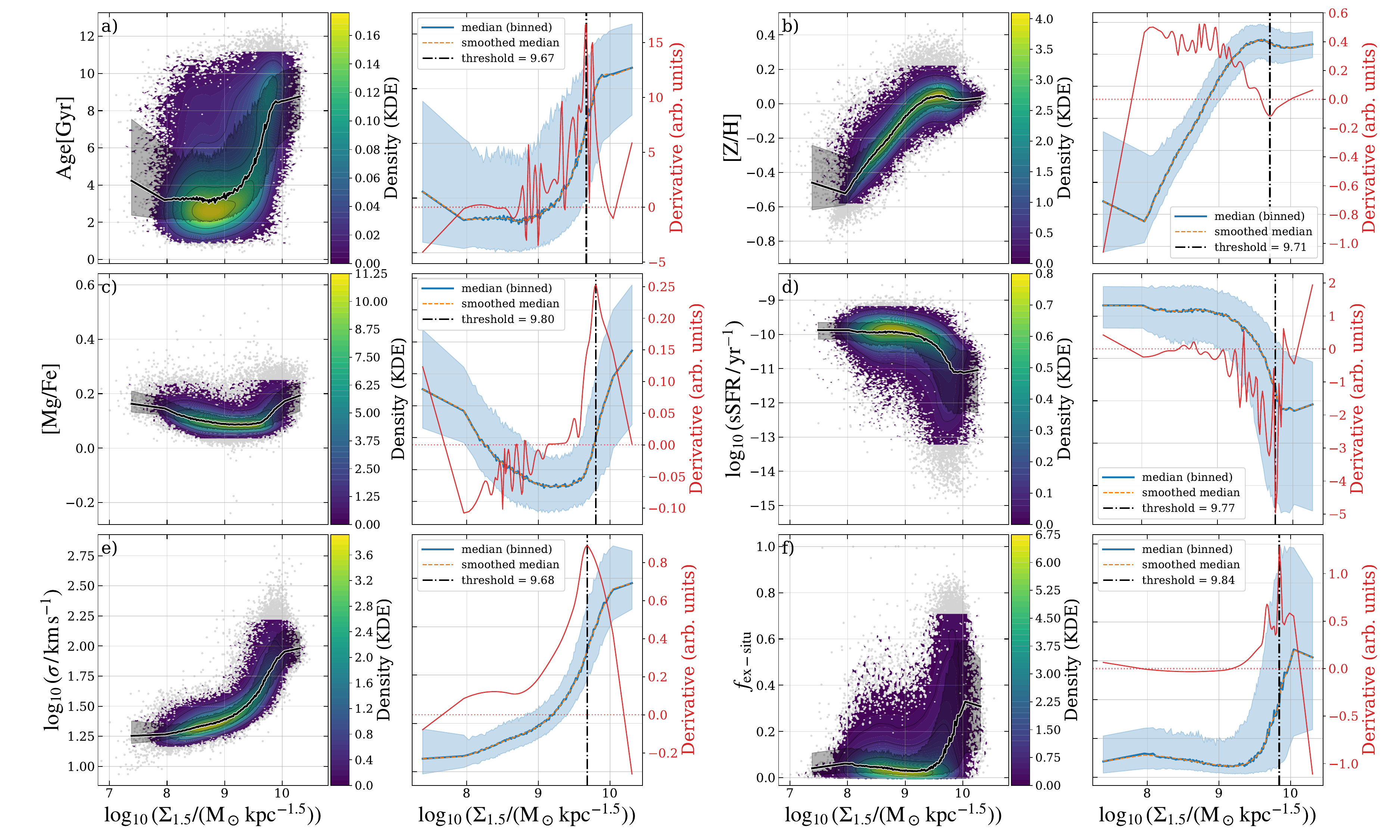}
\caption{Derivation of a physically motivated compactness threshold from the one-dimensional relations between 
$\log_{10}(\Sigma_{1.5}/(\rm{M}_\odot\,\mathrm{kpc}^{-1.5}))$ and different galaxy-evolution tracers: 
(a) luminosity-weighted mean stellar age; 
(b) stellar metallicity; 
(c) stellar [Mg/Fe]; 
(d) specific star formation rate; 
(e) stellar velocity dispersion; and 
(f) ex-situ stellar mass fraction. 
\maxz{For each tracer, the left subpanel shows all individual galaxies (grey points) overlaid on a kernel density estimate (KDE), with the median relation shown in black and the 16th–84th percentile range in grey.} The right subpanel shows the smoothed median together with its derivative (red; right axis); the dashed vertical line marks the high-compactness maximum of the derivative adopted as the compactness threshold.}
\label{fig:fourpanels_masssize_1d}

\end{figure*}

To quantify the trends identified in the stellar mass--size plane, we project each galaxy property onto a one-dimensional measure of structural compactness. We adopt the surface-density-like quantity

\begin{align}
\Sigma_{1.5} = \frac{M_\star}{R_\mathrm{e}^{1.5}},
\label{compactness}
\end{align}

\noindent where $M_\star$ is the stellar mass and $R_\mathrm{e}$ is the effective radius\footnote{\maxz{Throughout this paper, we use the stellar half-mass radius as the simulation analogue of the observational effective (half-light) radius employed by \citet{Barro13}.}}, following the definition introduced by \cite{Barro13}. 
This quantity was empirically motivated by the observation that quiescent galaxies follow a relation close to $R_\mathrm{e}\propto M_\star^{2/3}$, making $\Sigma_{1.5}$ an efficient discriminator between quiescent and star-forming systems. 

Other commonly used compactness indicators can be expressed within the same framework. For example, following the Zone of Exclusion one would obtain $\Sigma_{1.33} = M_\star/R_\mathrm{e}^{1.33}$, using the velocity dispersion $\sigma_\mathrm{e}$ as a compactness proxy would correspond to $\Sigma_{1.0} = M_\star/R_\mathrm{e}$ (assuming virial equilibrium and that stars dominate the mass), while using the effective stellar surface density would correspond to $\Sigma_{2.0} \equiv \Sigma_\mathrm{e} = M_\star/R_\mathrm{e}^2$. 
Adopting $\Sigma_{1.5}$ therefore provides a convenient compromise between these alternatives while also enabling direct comparison with observational studies, particularly those focused on compact and relic galaxy candidates (e.g. \citealt{Mills25}), which are predominantly quenched systems.

In the following, we use $\Sigma_{1.5}$ to test whether the visually identified transition in the stellar mass--size plane corresponds to a characteristic compactness threshold separating galaxy populations with distinct star-formation histories.

To reduce the impact of cross-correlations, we examine each quantity individually as a function of compactness, defined in Eq.~\ref{compactness}, for the full galaxy sample, as shown in the left-hand panels of Figure~\ref{fig:fourpanels_masssize_1d}. In each (left) panel, the black solid line represents the median trend, computed in bins of equal count, with 1,000 galaxies per bin, in order to ensure uniform statistical sampling along the compactness axis. The light grey band indicates the 16th--84th percentile range, while the colour-coding of the individual galaxies reflects the local number density estimated using a kernel density estimate (KDE).

The resulting relations reveal clear dependences on compactness for all quantities, although they are not all monotonic over the full compactness range. Within the range 
\mbox{$\log_{10}(\Sigma_{1.5}/(\rm{M}_\odot\,\mathrm{kpc}^{-1.5})) \approx 8-10$}, where most galaxies are concentrated, two broad regimes can be identified. In the low-compactness regime, 
$\log_{10}(\Sigma_{1.5}/(\rm{M}_\odot\,\mathrm{kpc}^{-1.5})) \lesssim 9.5$, the relations exhibit systematic trends whose strength and direction depend on the tracer considered. The trends are weaker and show larger scatter for sSFR, luminosity-weighted stellar age, and ex-situ mass fraction, while they are more structured for the other quantities. Toward higher compactness, however, the trends become more pronounced and coherent: luminosity-weighted stellar age, [Mg/Fe], stellar velocity dispersion, and ex-situ mass fraction all increase with compactness, while the sSFR decreases. Stellar metallicity shows a turnover, reaching higher values around the transition and weakly declining toward the most compact systems. 
This quantitatively supports a conceptual picture in which structural compactness encodes information about a galaxy's evolutionary history. 
Unlike the qualitative trends discussed in the previous section, this interpretation is now supported by a statistically robust 1D analysis. 

Interestingly, the ex-situ mass shows a significant scatter with a maximum of $\approx 0.7$ at compactness $\log_{10}(\Sigma_{1.5}/(\rm{M}_\odot\,\mathrm{kpc}^{-1.5})) \gtrsim 9.5$. 
This suggests that the compact population does not have a homogenous assembly history. 

By analogy with optical holography, we refer to this mapping as \textit{galactic morpholography}: the two-dimensional stellar mass--size plane effectively contains information about an implicit third, evolutionary time dimension. 
\maxz{The compactness transition itself, however, cannot be identified directly from the median compactness–tracer relations, since these often exhibit multiple local extrema arising from statistical fluctuations, residual population mixing, and the continuous nature of the galaxy distribution. We therefore define the characteristic compactness scale as the point where the smoothed median relation changes most rapidly within the high-compactness regime, corresponding to the maximum absolute first derivative. Rather than representing a sharp physical boundary, this threshold identifies the characteristic onset of the compact regime.}
Furthermore, we restrict this analysis to the high-compactness regime ($\log_{10}(\Sigma_{1.5}/(\rm{M}_\odot\,\mathrm{kpc}^{-1.5})) \gtrsim 9.5$), where the correlations are strongest and physically most relevant for the identification of relic systems. This ensures that the derived compactness threshold reflects the coherent evolutionary behaviour of compact galaxies, rather than being influenced by the weaker and more heterogeneous trends present at lower compactness.

The procedure is illustrated in the right-hand panels of Figure~\ref{fig:fourpanels_masssize_1d}. For each stellar population tracer, galaxies are first sorted by compactness and grouped into quantile bins targeting approximately 1000 galaxies each. In every bin, we compute the median of the tracer and the 16th and 84th percentiles. The median relation is then fitted with a weighted B-spline. \maxz{Because the bins are defined by quantiles rather than by enforcing exactly equal integer occupancies, the number of galaxies per bin differs slightly. We therefore assign each bin a weight $w_i=\sqrt{N_i}$ to account for these small residual differences in occupancy.} The derivative of the resulting spline is evaluated analytically, and the compactness threshold is identified from the strongest absolute derivative within a high-compactness search range. \maxz{To guard against selecting features caused by small-scale fluctuations in the smoothed relation, we define a robust derivative threshold from the median and median absolute deviation (MAD) of the absolute derivative within the search range. A candidate feature is accepted only if its amplitude exceeds this threshold. If no qualifying feature is found, the threshold is instead determined from the high-compactness end of the relation by scanning from high to low compactness and selecting the first bin whose derivative exceeds the robust threshold.}

 Applying this procedure independently to the luminosity-weighted stellar age, metallicity, [Mg/Fe], specific star formation rate, velocity dispersion and ex-situ mass fraction yields six individual threshold estimates. Their arithmetic mean defines the adopted compactness threshold,
\begin{align}
 \log_{10} \left(\frac{\Sigma_{1.5,\mathrm{crit}}}
 {\text{M}_\odot\,\mathrm{kpc}^{-1.5}}\right)
 = 9.75 \pm 0.10 .
\end{align}
\noindent The quoted uncertainty corresponds to the largest absolute deviation of the individual tracer thresholds from the adopted mean value, providing a conservative estimate of the systematic uncertainty associated with the threshold determination.

The remarkable similarity of the compactness thresholds obtained from the six independent tracers in Figure~\ref{fig:fourpanels_masssize_1d} supports the interpretation that the transition is not driven by a single observable, but reflects a coherent change in the evolutionary properties of compact galaxies.

\subsection{Heterogeneity of the compact galaxy population}

Strikingly, all examined galaxy-evolution tracers exhibit a change of behaviour within a similar compactness range. In this regime, higher compactness correlates with enhanced [Mg/Fe], older luminosity-weighted stellar ages, larger stellar velocity dispersions, and suppressed star formation activity. These stellar-population and dynamical properties consistently indicate systems that formed the bulk of their stars early and over short timescales, followed by rapid quenching; all of these being key characteristics of relic galaxies. 

Again, the results from the COLIBRE simulations are in excellent qualitative agreement with observational studies\footnote{Although we note that the compactness threshold usually defined in observations is higher \maxz{(e.g. 10.3 in \citep{Barro13})}, hence selecting only the most extreme tail of the distribution described here.}. \maxz{These studies likewise showed that the variation of galaxy properties, which as we saw is best described by diagonal lines in the mass--size plane, does not occur gradually as a function of compactness, or $\sigma_\text{e}$, or central density $\Sigma_1$, but instead exhibits a relatively sharp transition at a characteristic compactness scale (see fig.~2 of \citealt{Barro13}, or fig.~14 of \citealt{Cappellari23}).} A review of evidence for an empirical quenching transition is given by \citet[sec.~1]{ChenZhu2020}.

Taken alone, this coherent behaviour could suggest that high compactness is not only a necessary condition for relic galaxies, but a strong indicator of relic-like evolution. However, the distribution of ex-situ mass fractions complicates this interpretation. While the ex-situ fraction also shows a marked change of behaviour at high compactness (see Figure~\ref{fig:fourpanels_masssize_1d}), it points in a different direction: more compact galaxies often contain a large fraction of accreted stellar material. This may appear counter-intuitive, since accretion is often associated with size growth (e.g. \citealt{Naab+09}). Instead, it points to a more complex assembly history, in which some compact galaxies can experience significant accretion without losing their compact structure. Alternatively, some high-compactness systems with elevated ex-situ fractions may be stripped remnants, whose present-day compactness was enhanced by the tidal removal of their outer stellar envelopes rather than solely preserved since early formation. This behaviour is consistent with the stellar-mass gradient seen in the ex-situ mass-fraction panel of the mass--size plane (Figure~\ref{fig:fourpanels_masssize}f). Hence, both qualitatively and quantitatively, the evidence reinforces the conclusion that SCGs do not form a homogeneous class, a point that we explore further below. As such, compactness alone does not uniquely determine relicness, in line with recent observational evidence that CMGs are better described by a coupled structural--dynamical--evolutionary manifold than by compactness alone \citep{Spiniello26}. 

Finally, it is important to point out that the position of the compactness threshold directly affects the number of galaxies identified as relic candidates. Since compactness, defined according to Eq.~\ref{compactness}, is a continuous parameter, previous studies have often relied on empirical or percentile-based cuts to define CMGs. In contrast, the threshold derived here provides a physically motivated lower bound based on the underlying evolutionary behaviour, rather than an arbitrary selection.  Importantly, this definition does not restrict the sample to the most extreme tail of the compactness distribution, as in some observational studies \citep{Trujillo+09_superdense}, but instead captures a broader population located at the transition where compactness, quenching, and enhanced [Mg/Fe] become closely linked. 


\section{Simulated Ancient Galaxies (SAGs)}

\label{sec:relics}
To identify relic galaxies among the compact galaxy population, an additional quantity is required that captures the evolutionary aspect of relicness beyond morphology alone. Observational surveys such as \INSPIRE\ and \EINSPIRE\ 
introduced the `Degree of Relicness' (DoR), a dimensionless indicator designed to quantify how closely a galaxy resembles an untouched relic system in terms of \chiara{its star formation history}. 

A quantitative definition of the DoR was first given in \citet{Spiniello+21}, following the idea already introduced in \citet{Ferre-Mateu+17}. Specifically, the DoR is defined as: 

\begin{align}
\label{DoR}
\text{DoR} =
\frac{1}{3}
\bigg[
f_{M^*_{z=2}}
+
\frac{0.5 \, \mathrm{Gyr}}{t_{75}}
+
\frac{0.7 \, \mathrm{Gyr} + (t_{\mathrm{Uni}} - t_{\mathrm{fin}})}
{t_{\mathrm{Uni}}}
\bigg],
\end{align}
\noindent 
\chiara{where $f^{M_\star}_{z=2}$ denotes the fraction of the final formed stellar mass (the initial masses of stars belonging to the galaxy
at the observation epoch) that had formed by $z=2$, regardless of
which progenitor hosted those stars at the time. The quantity $t_{75}$ is the cosmic time by which 75\% of that mass had formed, $t_{\rm fin}$ is a corresponding late-time percentile of the cumulative star-formation history, and $t_{\rm Uni}$ is the age of the Universe at the observation epoch. In the observational DoR definition, $t_{\rm fin}$ was termed the ``final assembly time'' and corresponds to the epoch at which the cumulative star-formation history reaches 99.8\% of its final formed stellar mass. Operationally, it records when the stars present at the observation epoch formed, irrespective of their birth sites and it does not determine when ex-situ stars joined the main progenitor.} 
However, in adapting the same definition to the COLIBRE simulation, we slightly amended the definition and used the epoch by which 95\% of the final formed stellar mass is in place, in order to account for the finite numerical resolution of the simulation. 
At low stellar masses, the original observational definition becomes highly sensitive to small numbers of late-forming star particles, causing many galaxies to be assigned artificially late $t_{\rm fin}$ values. The adoption of a 95\% threshold reduces this numerical sensitivity and produces a smoother distribution across the stellar mass--size plane while preserving the qualitative behaviour of the DoR (Appendix~\ref{app:tfin}). 

According to the above definition, \maxz{the DoR usually ranges from 0 to 1.} Larger DoR (DoR $>0.6$--$0.7$) values correspond to galaxies that formed the great majority of their stellar mass at early times and over short timescales, i.e. relics, whereas galaxies with prolonged or ongoing star formation histories tend towards $\text{DoR}\approx0$. For the COLIBRE simulated galaxies described in Section~\ref{sec:colibre}, the quantities entering the DoR are derived from the cumulative stellar mass growth reconstructed from the SFHs. \maxz{Consequently, all stellar masses entering the DoR correspond to cumulative formed stellar masses reconstructed from the SFHs, rather than present-day surviving stellar masses.} In particular, we determine the epochs at which fixed fractions of the final stellar mass have formed and the fraction of final formed stellar mass that had formed by $z=2$.

In observational studies, the DoR has so far been applied exclusively to preselected CMGs, where compactness is already imposed as an implicit condition. Within this context, galaxies with DoR values close to unity correspond to extreme relics that formed nearly all of their stellar mass early and over short timescales, while lower values indicate more extended formation histories. 
In the present simulation analysis, by contrast, we apply the DoR to the full galaxy population independently of morphology. \maxz{As a consequence, high DoR values do not uniquely identify relic galaxies, because early \chiara{star formation} 
alone does not guarantee that a galaxy remains compact until the present day. Subsequent structural evolution -- for example through dry mergers that increase galaxy size while adding predominantly old ex-situ stars, or through evolutionary pathways that maintain extended morphologies despite early star formation -- can produce systems with high DoR values that no longer satisfy the structural definition of relics.}

To distinguish this evolutionary aspect from the structural definition of relics, we therefore refer to galaxies with high DoR values as simulated ancient galaxies (SAGs). In the following, we adopt a threshold of $\text{DoR} > 0.6$, motivated by the observational results presented in \citet{Spiniello+24} and \citet{Mills25}. This threshold is not intended to represent a sharp physical boundary, but rather aims to isolate galaxies with strongly enhanced relic-like star-formation histories while retaining a statistically meaningful sample size of 490 SAGs with $M_\star \geq 10^9 \, \rm{M}_\odot$. 
Relic candidates are then identified as the overlap between SAGs and SCGs, ensuring that both the evolutionary and morphological aspects of relicness are taken into account simultaneously. 

\begin{figure*}
    \centering    \includegraphics[width=1.05\textwidth, trim={1cm 0cm 0cm 0cm}]{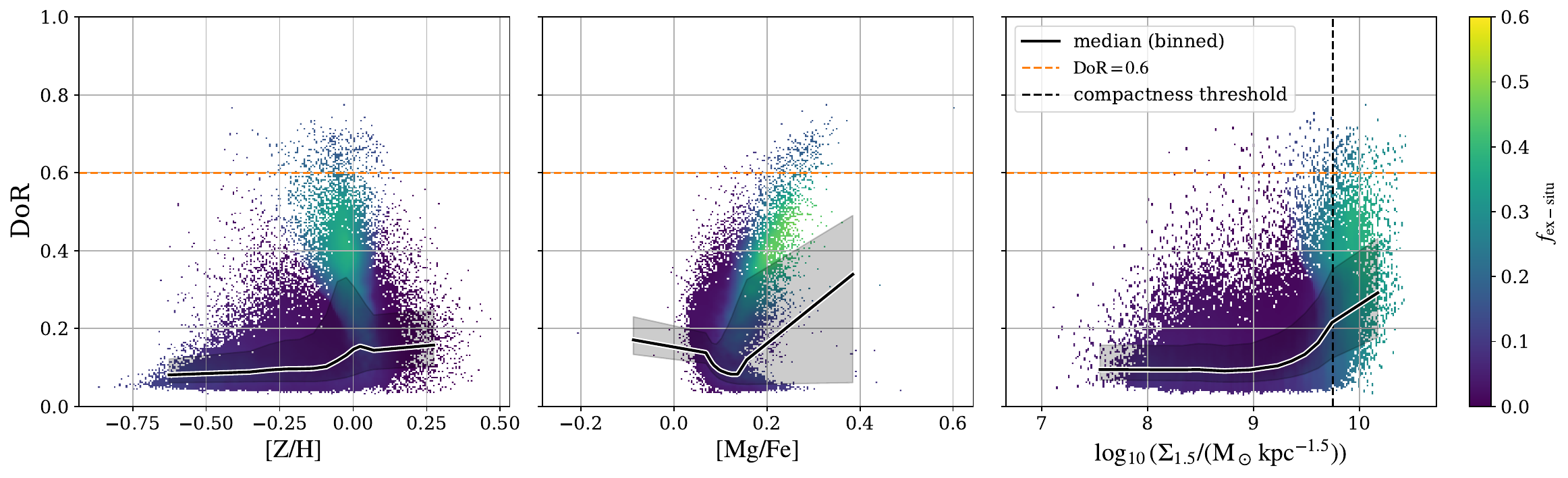}
    \caption{DoR plotted against stellar-population and structural diagnostics \maxz{for all COLIBRE galaxies at $z=0$ with stellar masses $M_\star \geq 10^9\,{\rm M_\odot}$}, coloured by the LOESS-smoothed ex-situ mass fraction: stellar metallicity (left); stellar [Mg/Fe]
(middle); compactness (right). LOESS is applied only to the colour coding; galaxy coordinates and the median and percentile trends are computed from the original, unsmoothed data.}
    \label{fig:3panel_DoRvsquantity}
\end{figure*}

\begin{figure*}
    \centering   \includegraphics[width=1.0\textwidth]{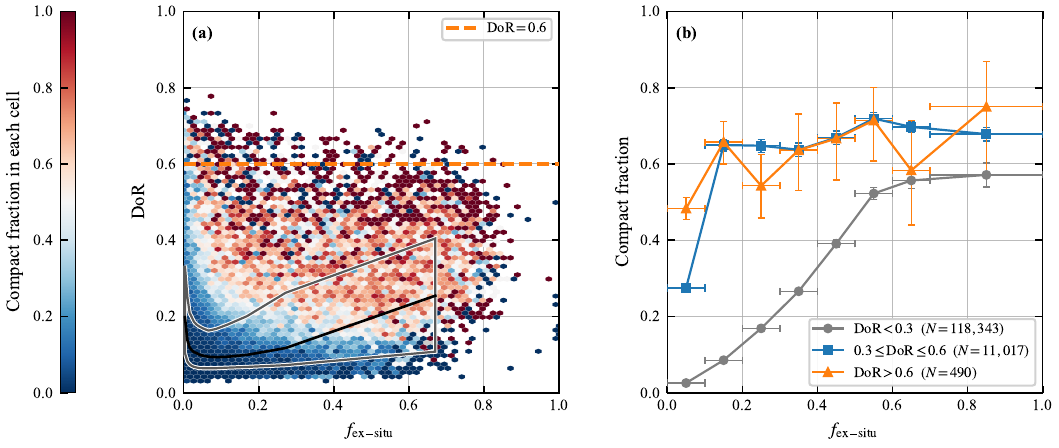}
    \caption{DoR and ex-situ mass fraction for COLIBRE galaxies at $z=0$
with $M_\star \geq 10^9\,{\rm M_\odot}$.
\emph{Left:} Colours show the fraction of compact galaxies ($\log_{10}(\Sigma_{1.5}/({\rm M_\odot\,kpc^{-1.5}}))>9.75$) within each
hexagonal cell, ranging from zero (blue) to unity (dark red).
The black line shows the median DoR, and the grey envelope indicates
the 16th--84th percentile range. The orange dashed line marks ${\rm DoR}=0.6$.
\emph{Right:} Compact fraction as a function of ex-situ mass fraction
for galaxies with ${\rm DoR}<0.3$ (grey),
$0.3\leq{\rm DoR}\leq0.6$ (blue), and ${\rm DoR}>0.6$ (orange). 
All three groups use identical bins, with the final bin spanning
$0.7\leq f_{\rm ex\text{-}situ}\leq1$.
Horizontal bars indicate bin extents while vertical bars show approximately
68\% binomial confidence intervals. 
Of the 490 ancient galaxies (${\rm DoR}>0.6$), 298 (60.8\%) have
$f_{\rm ex\text{-}situ}<0.1$.}
    \label{fig:panel-DoRvsexsitu}
  \end{figure*}

\begin{figure*}
\centering
\includegraphics[width=1.0\textwidth]{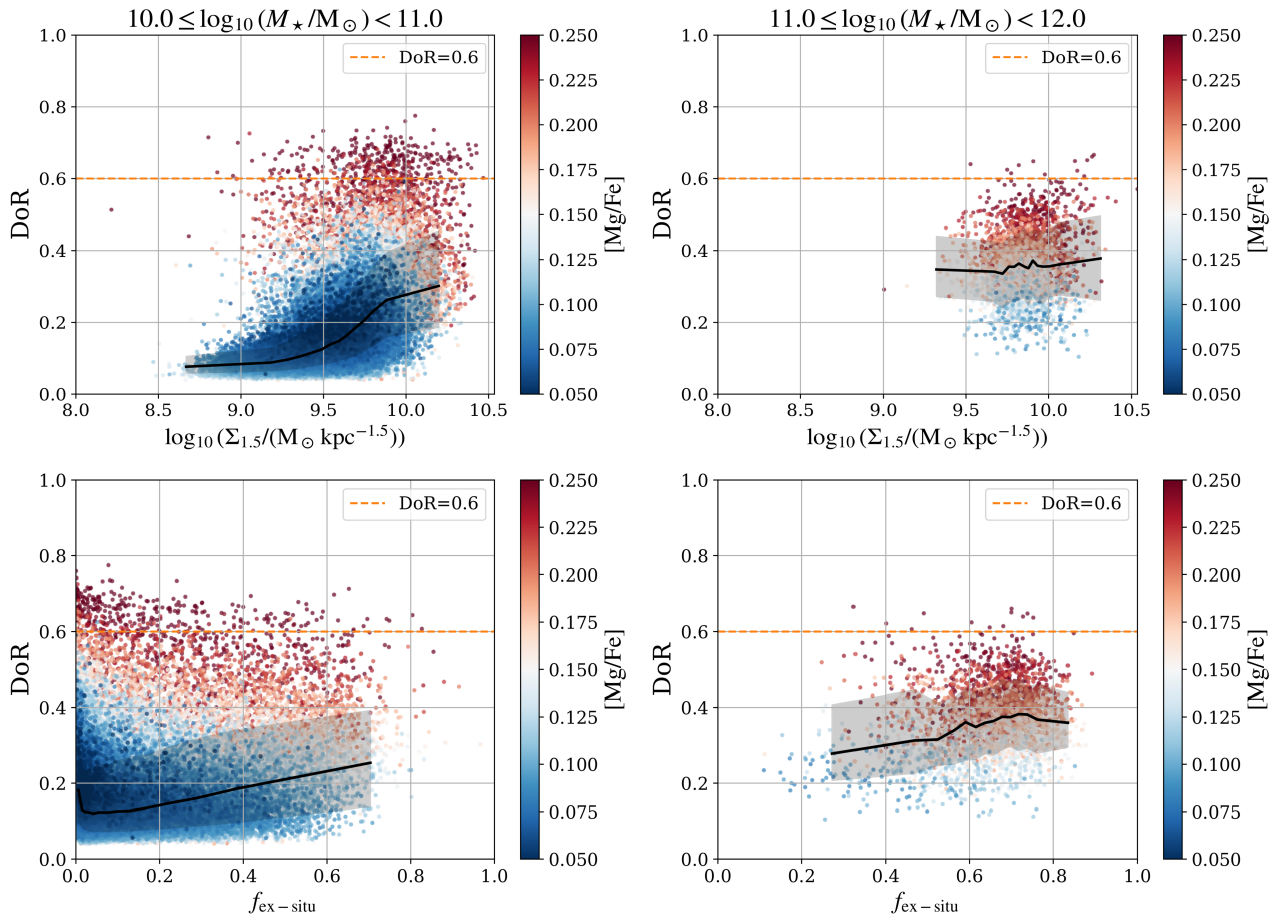}




\caption{ 
DoR plotted against compactness (upper row) and ex-situ mass fraction (lower row) for the intermediate $(10^{10} - 10^{11} \, \rm{M}_\odot$, left) and the high-mass bin $(10^{11} - 10^{12} \, \rm{M}_\odot$, right). Galaxies are coloured by stellar [Mg/Fe]. Point colours encode stellar [Mg/Fe] on the same scale in all
panels. Black curves show the median DoR in bins of the quantity
on the horizontal axis; grey bands show the 16th--84th percentile
ranges. }
\label{fig:DoR_vs_quant_bin}

\end{figure*}

\subsection{The relation between compactness and relicness}


\chiara{Across the stellar mass--size plane, the DoR follows a qualitatively
similar distribution to the luminosity-weighted mean stellar age
(see Figure~\ref{fig:fourpanels_masssize}a). This is expected from its
construction, since both quantities trace early stellar mass assembly.
The crucial question, however, is whether the DoR also correlates with
the additional quantities identified in Section~\ref{sec:SCGs} as
signatures of relic-like evolution, in particular structural compactness.
To address this, we analyse the relation between DoR and the previously
introduced evolutionary tracers, as shown in
Figure~\ref{fig:3panel_DoRvsquantity}. In analogy to the compactness
analysis, we compute median trends and 16th--84th percentile ranges. The galaxy distribution is colour-coded by the LOESS-smoothed
ex-situ mass fraction.  
Unlike DoR, which primarily traces the timing and duration of star
formation, the ex-situ fraction probes the cumulative importance
of merger-driven assembly by $z=0$. } It therefore provides complementary information on the second evolutionary phase of galaxy growth, allowing us to test whether galaxies with similar DoR values also share comparable assembly histories. Since the ex-situ fraction can be measured directly from the simulation, it also offers a useful link between the observable stellar population properties and the underlying formation pathways. We do not include stellar age in this figure because it is intrinsically connected to the DoR by construction. Instead, we focus on stellar metallicity, [Mg/Fe], and compactness, which trace chemical enrichment, star formation timescales, and structural evolution, respectively.  The remaining quantities \maxz{(i.e. sSFR and stellar velocity dispersion)} do not provide additional information beyond the trends discussed here and are therefore not explicitly shown.

The resulting correlations reveal a coherent evolutionary picture. Increasing DoR is associated with enhanced [Mg/Fe], supporting the interpretation that high-DoR systems experienced short star formation timescales and early chemical enrichment. 

The relation with stellar metallicity is instead less monotonic. The highest-DoR objects are generally metal-rich, but they are not necessarily the most metal-rich systems in the sample. In particular, the left panel of Figure~\ref{fig:3panel_DoRvsquantity} shows that DoR peaks around $-0.2 \lesssim [\text{Z/H}] \lesssim 0.2$. This suggests that metallicity alone is not a monotonic tracer of relicness, even though high-DoR systems preferentially occupy a relatively metal-rich regime. 

Most importantly, Figure~\ref{fig:3panel_DoRvsquantity} demonstrates that the DoR also correlates strongly with structural compactness. Strikingly, the compactness threshold derived in Section~\ref{sec:comp_crit} \maxz{matches the characteristic transition in the median DoR–compactness relation}, independently reinforcing the physical relevance of the compactness threshold. This agreement demonstrates that compactness and early \maxz{stellar populations} are deeply connected, supporting the morpholography picture introduced earlier. 
At the same time, the relation between compactness and DoR also highlights an important distinction between compactness and relicness. Although high DoR values preferentially occur among compact galaxies, a substantial population of ancient galaxies (with DoR $>0.6$) remains below the compactness threshold. 
These systems exhibit early stellar mass formation but lack the compact structure expected for relic galaxies, albeit still clustering slightly below the $\log_{10}(\Sigma_{1.5}/(\rm{M}_\odot\,\mathrm{kpc}^{-1.5})) = 9.75$ value. 4.9\% of the SAGs are found at $\log_{10}(\Sigma_{1.5}/(\rm{M}_\odot\,\mathrm{kpc}^{-1.5}))< 9$. Conversely, only $\sim2$\% of the compact galaxies have DoR values of $\text{DoR} > 0.6$. In fact, although the median DoR value increases at higher $\log_{10} \Sigma_{1.5}$, the maximum median reached by the compact galaxies is 
$\approx 0.3$. Consequently, neither compactness nor DoR alone is sufficient to identify relic systems in the full simulated galaxy population. Relic candidates must instead be selected by combining the structural compactness criterion with the evolutionary information encoded in the DoR.

\subsection{Ex-situ mass fraction and subpopulations}
The ex-situ mass fraction provides further evidence that ancient
galaxies do not form a homogeneous evolutionary class. The left panel
of Figure~\ref{fig:panel-DoRvsexsitu} shows a population of 298 galaxies
(60.8\% of all SAGs) with $\text{DoR}>0.6$ at very low ex-situ mass
fractions ($f_{\rm ex\text{-}situ}<0.1$), consistent with systems that
assembled almost all of their stellar mass early and in situ.
These objects have assembly histories most consistent with the
expected formation history of classical relic galaxies, having
assembled almost all of their stellar mass early and experienced
little subsequent stellar accretion. However, only those that also
satisfy the compactness criterion qualify as relic candidates. At the same time, the median DoR of the full population increases gradually towards larger ex-situ mass fractions, indicating that ancient \maxz{star formation} histories are not restricted to merger-free systems.

\chiara{The right panel of Figure~\ref{fig:panel-DoRvsexsitu}
uses the low- ($\mathrm{DoR}<0.3$), intermediate-
($0.3\leq\mathrm{DoR}\leq0.6$), and high-DoR
($\mathrm{DoR}>0.6$) groups adopted in E-INSPIRE
\citep{Mills25}. The division at $\mathrm{DoR}=0.3$ is used here
for comparison, whereas $\mathrm{DoR}>0.6$ defines SAGs.}
At low ex-situ fractions, the compact fraction rises markedly
from the low- to the high-DoR group. At larger ex-situ fractions,
even low-DoR galaxies frequently have compact structures.
Hence, compact systems populate the entire
$f_{\rm ex\text{-}situ}$ range as well as the entire DoR range,
from extended or more prolonged star-formation histories to
genuinely ancient ones. Compactness therefore appears to be a
necessary but not sufficient condition for relic-like evolution.
This behaviour reinforces the distinction established in the
previous sections between compact galaxies and ancient stellar
systems, and demonstrates that the two populations overlap only
partially to give rise to relic candidates in the simulation. 

The coexistence of high-DoR galaxies at both low and moderate-to-high ex-situ mass fractions further shows that early stellar formation does not uniquely determine subsequent assembly history. A galaxy can host ancient stellar populations and still have a large ex-situ fraction if a significant fraction of its stars formed early in external progenitors and were later accreted. Such systems are ancient in terms of their stellar populations, but they are not necessarily relics in the sense of having avoided the merger-driven growth phase. This points toward at least two different evolutionary pathways: a classical relic-like channel dominated by early in-situ formation with minimal subsequent accretion, and a merger-built ancient channel in which old stellar populations are assembled through substantial ex-situ growth. Stripping of a previously larger SAG might be a third channel.

We remind the reader that the ex-situ fraction adopted here integrates the full assembly history of each galaxy and therefore does not distinguish between mergers occurring before and after $z=2$. 
When we recompute this quantity by considering only ex-situ stellar mass accreted at $z\leq2$, many high-DoR galaxies show a substantial reduction in their ex-situ fraction, in some cases from values above 0.5 to below 0.2. Nevertheless, a small subset of high-DoR galaxies ($\approx 12\, \%$) retains a high ex-situ mass fraction even under this restricted definition. 

This dichotomy between the two populations of ancient galaxies in COLIBRE becomes even clearer when the analysis is carried out in 
different stellar mass ranges, as shown in Figure~\ref{fig:DoR_vs_quant_bin}. We focus this comparison on $10^{10}\leq M_\star/{\rm M}_\odot<10^{12}$, which contains most
compact, high-DoR systems in our sample. We split at $10^{11}\,{\rm M}_\odot$ to contrast the intermediate-mass population with the more massive regime, where ex-situ fractions are typically higher (Figure~\ref{fig:fourpanels_masssize}f). However, we stress that this split does not
imply a sharp physical boundary. At intermediate masses ($10.0 \leq \log_{10}(M_\star/M_{\odot}) < 11.0$), SAGs represent $0.78\%$ of the population and preferentially occupy the low ex-situ regime (left panels), \maxz{consistent with the theoretical expectation that relic galaxies should experience little merger-driven growth after their early formation.}  

At the high-mass end ($11.0 < \log_{10}(M_\star/M_{\odot}) < 12.0$, SAGs represent $0.45\%$ of the  population. Here galaxies have preferentially intermediate DoR values while simultaneously exhibiting systematically enhanced ex-situ mass fractions (bottom right panel in Figure~\ref{fig:DoR_vs_quant_bin}). A similar \maxz{loss of clear correlation} appears in the compactness--DoR relation: the strong compactness dependence of DoR observed at intermediate masses largely disappears toward the highest stellar masses.

For both quantities, there is a clear separation of galaxies with enhanced [Mg/Fe] around the SAG threshold and lower values for the low DoR end across both mass ranges, confirming that a short period of star formation traced by [Mg/Fe] is a necessary requirement for relic systems to form. Generally, no such clear correlation can be found within the 50 kpc aperture when colouring by stellar metallicity. However, decreasing the aperture reveals metal-rich cores with a stronger metallicity–compactness correlation, while no comparable trend emerges with ex-situ mass fraction. This suggests that some high-DoR galaxies may host enhanced metallicities in their compact central regions, as expected for relic systems, whereas the inclusion of stars at larger radii dilutes this signal. One possible explanation is that subsequent accretion of a more metal-poor outer envelope reduces the average metallicity measured within larger apertures.

Taken together, these trends indicate that the connection between compactness, ancient DoR and true relicness-like features weakens toward the massive end of the galaxy population. 
Massive high-DoR galaxies appear to follow a distinct evolutionary pathway characterised by old stellar populations, continued accretion, and enhanced ex-situ growth. This interpretation is consistent with the horizontal ex-situ gradient observed across the stellar mass--size plane (see Figure \ref{fig:fourpanels_masssize}f) and suggests that ancient galaxies, rather than compact galaxies as a whole, separate into at least two physically distinct assembly channels: a compact, low-ex-situ channel closer to the classical relic picture, and a more massive, merger-built channel with larger ex-situ fractions. The latter population does not fit in the classical phenomenological picture of relic galaxies. A detailed analysis of merger histories, environmental processing, and possible tidal stripping will be required to establish the physical origin of these channels, and will be presented in a follow-up paper (Zemsch et al., in prep.).

\begin{figure}
    \centering    \includegraphics[width=0.5\textwidth,
    trim={0.2cm 0cm 0cm 0cm}]{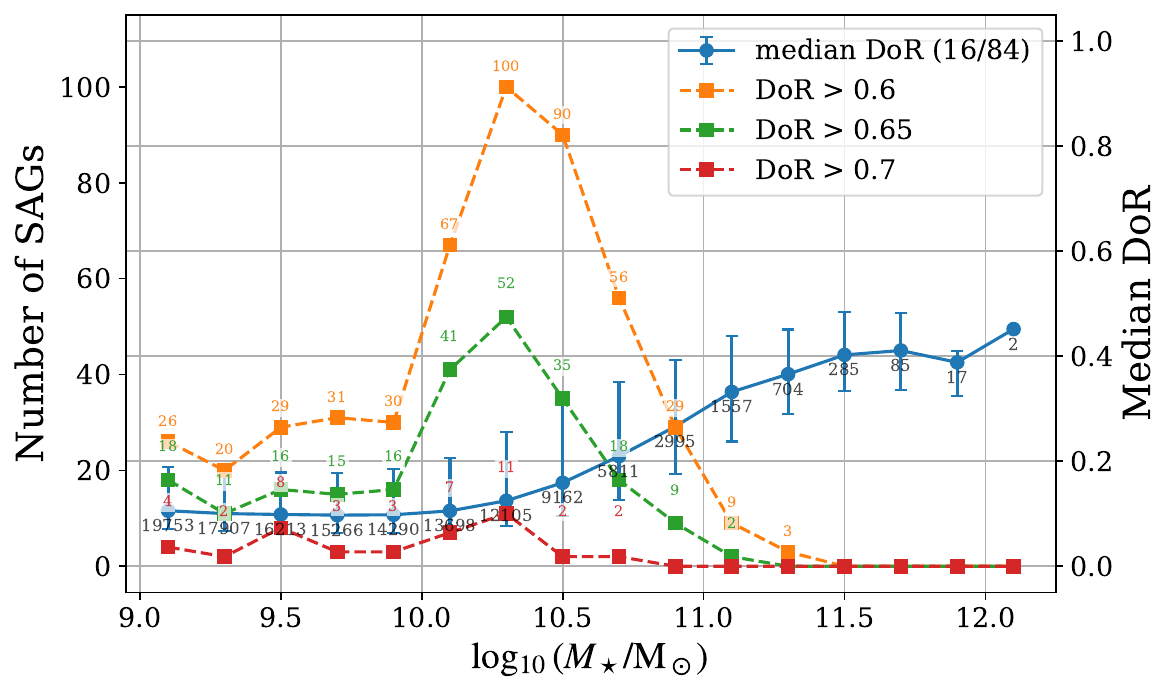}
\caption{Number of simulated ancient galaxies per $0.2$-dex stellar-mass bin for different DoR thresholds. As a reference, the median DoR of all galaxies in each mass bin is plotted in blue. The respective galaxy counts are denoted above the datapoints.}
\label{fig:SAG-counts}
\end{figure}

The stellar mass dependence is further illustrated in Figure~\ref{fig:SAG-counts}, which compares the number of SAGs selected at different DoR thresholds with the median DoR (blue solid line) of the full galaxy population in each $0.2$-dex stellar-mass bin.  
While the median DoR increases systematically toward higher stellar masses, the number of galaxies exceeding a fixed high-DoR threshold does not simply follow this median trend. Instead, the SAG counts peak at intermediate masses, around $10.1 \lesssim \log(M_\star/\rm{M}_\odot) \lesssim 10.7$, and decline toward higher stellar masses. This illustrates that the high-mass population can have relatively large average DoR values without producing the largest number of extreme high-DoR outliers. 
\maxz{We note that the absolute number of galaxies naturally decreases at the high-mass end, including the number of SAGs. However, the fraction of galaxies classified as SAGs likewise has a broad maximum of about $1\%$ over $10.4\lesssim\log_{10}(M_\star/\rm M_\odot)<11.0$ 
and decreases toward higher stellar masses ($0.58\%$ at $11.0\leq\log_{10}(M_\star/\rm M_\odot)<11.2$), indicating that the trend is not solely driven by the underlying stellar mass function.} 
In this sense, the most ancient systems selected by a fixed DoR threshold preferentially occupy an intermediate-mass regime, while the highest-mass galaxies are more typically characterised by elevated median DoR and larger ex-situ fractions. The persistence of this behaviour for different DoR thresholds indicates that it is not driven by the exact threshold choice, but reflects the underlying mass dependence of early stellar formation in the simulated galaxy population. 


\section{Simulated Relic Galaxies (SRGs)}
\label{sec:compact-relic-relation}

Following the arguments developed above, we identify simulated
relic candidates (SRGs) as the overlap between SCGs and SAGs. In other words, relic candidates are required to satisfy both the compactness threshold $\rm  log_{10}(\Sigma_{\rm 1.5,crit}/(M_{\odot}\,kpc^{-1.5}))=9.75$ derived in Section~\ref{sec:comp_crit} and the evolutionary criterion $\text{DoR}>0.6$. \chiara{This operational selection identifies galaxies with compact
present-day structures and early star-formation histories, but
does not impose a limit on merger-driven growth after $z=2$.} 

\begin{figure}
    \centering   \includegraphics[width=0.5\textwidth,
    trim={0.2cm 0cm 0cm 0cm}]{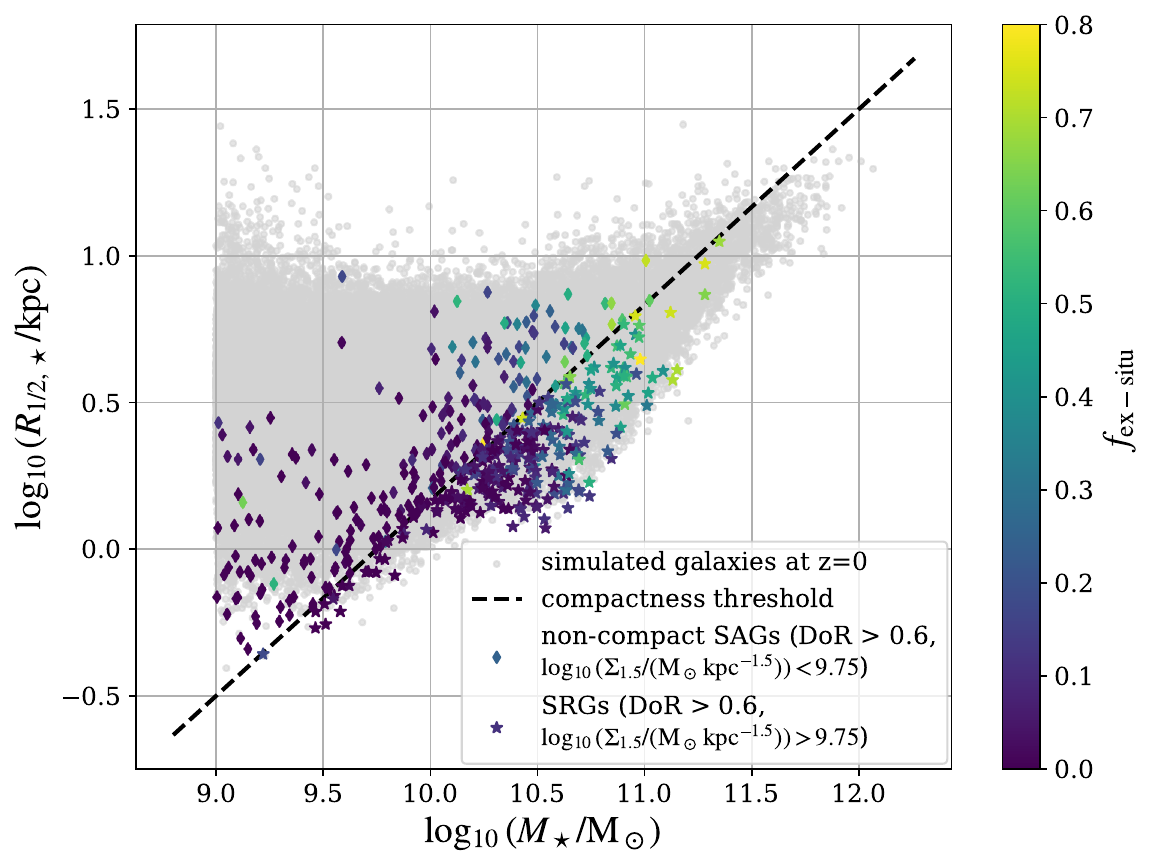}
    \caption{Simulated ancient galaxies colour-coded by ex-situ mass fraction in the stellar mass-size plane. \maxz{The dashed line corresponds to the compactness threshold $\rm  log_{10}(\Sigma_{\rm 1.5,crit}/(M_{\odot}\,kpc^{-1.5}))=9.75$. SAGs exceeding this threshold, i.e. SRGs, are shown as star symbols, while the remaining SAGs are shown as diamonds. The low DoR galaxies of the sample are depicted in grey.}}
    \label{fig:relicsample}
\end{figure}

The combined selection yields 267 SRGs at $z=0$. \maxz{The present-day morphologies of a representative subset of these SRGs are shown in Appendix~\ref{app:morphology}, including examples of both relatively isolated systems and galaxies with nearby companions or surrounding structures indicative of future interactions.} Their distribution
in the stellar mass--size plane is shown in Figure~\ref{fig:relicsample}, colour-coded by ex-situ stellar mass fraction. A clear pattern emerges: most SRGs occupy an intermediate stellar-mass range, $10.0 \lesssim \log_{10}(M_\star/\rm{M}_\odot) \lesssim 10.7$, where they exhibit both high compactness and comparatively low ex-situ mass fractions. This is broadly consistent with the classical picture of relic galaxies as systems that assembled the bulk of their stellar mass rapidly at early times. However, the integrated ex-situ fraction alone does not indicate when mergers occurred, and therefore cannot by itself distinguish between galaxies that experienced significant accretion before quenching and those that assembled substantial stellar mass during the later merger-driven growth phase. 

Towards higher stellar masses, however, the properties of the SRG population change systematically. Even among galaxies satisfying both the compactness and DoR criteria, the ex-situ stellar mass fraction increases substantially. Ex-situ fractions exceeding $\approx0.5$ imply that a significant fraction of the present-day stellar mass was assembled through mergers, which appears difficult to reconcile with the traditional picture of relic galaxies as systems that largely avoided the second, accretion-driven phase of galaxy evolution.
However, we note that the ex-situ fraction adopted throughout most of this work integrates the full assembly history of each galaxy and therefore does not distinguish between accretion occurring before and after $z=2$. When we instead calculate the ex-situ fraction using only stellar mass accreted at $z<2$, hence splitting between the first and second phase, as assumed in observations, the massive SRG population displays a substantial reduction in ex-situ fraction. This indicates that a significant part of its merger-driven assembly occurred at or before the main quenching epoch around $z=2$, rather than during the later evolutionary phase normally associated with the size growth of massive galaxies. Nevertheless, a subset of massive SRGs ($\approx 15\, \%$) retains high ex-situ fractions even when only accretion at $z<2$ is considered. These objects therefore occupy a transitional regime between classical relics and compact ancient galaxies that experienced significant later assembly. 

\begin{figure}
    \centering    \includegraphics[width=0.5\textwidth,
    trim={0.2cm 0cm 0cm 0cm}]{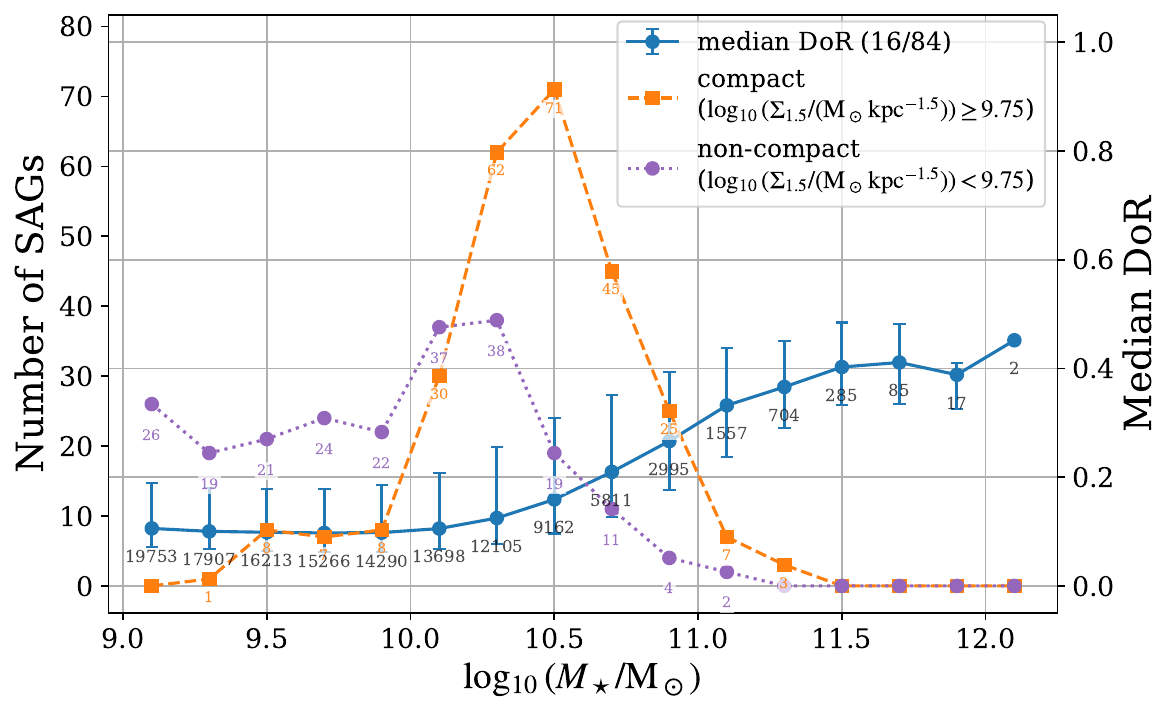}
    \caption{Distribution of SAGs (i.e. galaxies with DoR > 0.6) across different $0.2$-dex stellar-mass bins, split by compactness threshold (i.e. $\log_{10}(\Sigma_{1.5}/(\rm{M}_\odot\,\mathrm{kpc}^{-1.5}))=9.75$). As a reference, the median DoR of all galaxies in each mass bin is plotted in blue. The respective galaxy counts are denoted above the data points.}
    \label{fig:SAG-counts(split)}
\end{figure}

The stellar-mass dependence of the ancient population is further illustrated in Figure~\ref{fig:SAG-counts(split)}, which compares the distributions of compact and non-compact SAGs. Compact SAGs, i.e. SRGs, show a pronounced peak at intermediate stellar masses, $\approx10^{10.5}\,\rm \rm{M}_\odot$, whereas non-compact SAGs are distributed more broadly and peak at somewhat lower stellar masses, around $10^{10}$--$10^{10.5}\,\rm \rm{M}_\odot$.
\chiara{These counts are modulated by the declining total number of galaxies
per mass bin. Indeed, the fraction of all galaxies classified as SRGs
remains approximately $0.8\%$ throughout
$10.4\leq\log_{10}(M_\star/\rm M_\odot)<11.0$. The count maximum near
$10^{10.5}\,{\rm M}_\odot$ therefore does not imply a narrow peak
in SRG incidence.}


\subsection{Black hole masses and AGN-driven quenching}

\begin{figure}
    \centering
    \includegraphics[width=\columnwidth,
    trim={0.2cm 0cm 0cm 0cm}]{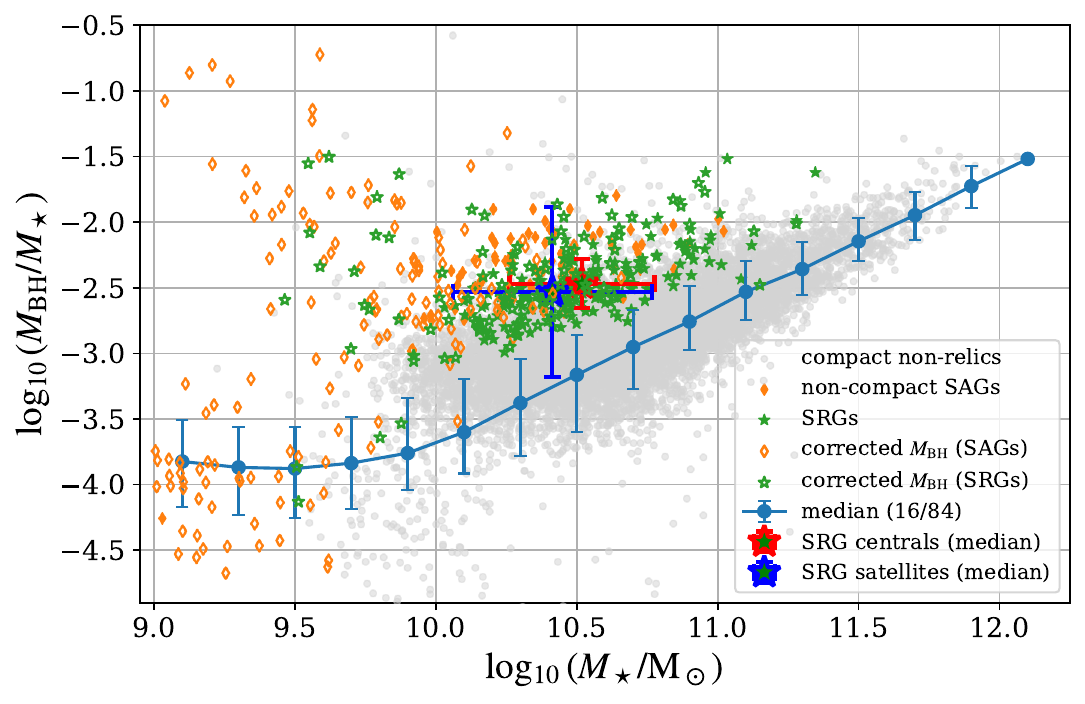}
    \caption{Black hole-to-stellar mass ratio as a function of stellar mass for relic candidates (SRGs; green stars), non-compact SAGs (orange diamonds), compact non-relics (grey points), and the full galaxy population. The blue line and error bars show the median relation and its 16th--84th percentile range for the full population. Open symbols indicate galaxies whose black-hole masses are affected by the repositioning issue and should therefore be regarded as lower limits. The median locations of central and satellite SRGs are outlined in red and blue, respectively, with error bars indicating the standard deviation in both directions.}
    \label{fig:BHmass}
\end{figure}

A natural question is whether the rapid quenching required to form relic systems can be linked to a dominant physical mechanism. Recent COLIBRE studies have shown that the observed abundance of massive quiescent galaxies at high redshift ($z>2$) can be reproduced reasonably well within a standard $\Lambda$CDM framework, without requiring 
unusually high star-formation efficiencies \citep{Chaikin+25,Chandro-Gomez+26a}. \maxz{A further important ingredient is the allowance for super-Eddington black hole accretion, which accelerates early black hole growth and the onset of AGN feedback in COLIBRE \citep{Chaikin+26c}.} 
Building on this, \citet{Chandro-Gomez+26b} demonstrated that AGN feedback is the primary mechanism responsible for quenching massive galaxies at high redshift. 
Quenched systems host more massive black holes and experience stronger cumulative AGN energy injection than their star-forming counterparts. Their progenitors also assemble preferentially in overdense environments before quenching, suggesting that dense surroundings promote the gas inflows and black-hole growth that eventually trigger AGN-driven quenching. 
Consistent with this picture, the thermal COLIBRE AGN model quenches high-redshift massive quiescent galaxies more efficiently than the hybrid thermal+jet model, because black holes grow more slowly in the hybrid case and jet feedback operates on longer timescales \citep{Chandro-Gomez+26b}. 

%
These early massive quenched galaxies represent a plausible progenitor population for present-day relic galaxies. If relics originate from the most extreme early-forming and rapidly quenched systems, signatures of the black-hole growth associated with that quenching phase may still be visible at $z=0$. More generally, black hole masses have been found to be the main predictor of AGN feedback quenching galaxies in both observations and simulations (e.g. \citealt{Piotrowska+22}) in the local Universe. Observationally, overmassive black holes have also been reported in several nearby relic galaxies \citep{van_den_Bosch_2012,Ferre-Mateu+15,Walsh+15,Scharwachter16}.
We therefore examine whether simulated relic candidates occupy a distinct region of the black hole-to-stellar mass relation relative to the general galaxy population. 

Figure~\ref{fig:BHmass} shows the black hole-to-stellar mass relation across the full galaxy sample, together with the locations of SAGs and SRGs. Both relic candidates and, more generally, SAGs preferentially host overmassive black holes relative to typical galaxies of the same stellar mass, with offsets of $\approx 0.25$--$2.5$ dex above the median relation. This indicates a close association between black-hole growth and the early stellar formation histories identified in the previous sections.
At lower stellar masses, some SAGs lie below the median relation because their black-hole masses are affected by the repositioning issue described in Section~\ref{sec:colibre}. These galaxies are shown with open symbols in Figure~\ref{fig:BHmass} and must be interpreted with caution. Because the reconstructed black-hole masses are lower limits, the corresponding points are expected to move upwards if their true black-hole masses at $z=0$ were known. In total, 254 out of 490 SAGs, including 107 out of 267 SRGs, are affected by this issue, the majority of which are satellites.

The presence of overmassive black holes in relic candidates suggests that a black-hole excess established during the initial formation phase can remain preserved in systems that subsequently experience limited stellar-mass growth. The enhanced black-hole masses of the SRGs are therefore qualitatively consistent with a scenario in which rapid early stellar assembly is accompanied by efficient black-hole growth and strong AGN feedback. In such a picture, substantial fractions of both the stellar and black-hole mass are assembled during a short, intense formation phase at high redshift. The resulting AGN activity may then contribute to the rapid suppression of star formation, leaving behind a compact, quenched galaxy with an elevated black hole-to-stellar mass ratio. \maxz{Tracing the stellar masses back to $z=2$ shows that they evolve only modestly. Even among the galaxies with the highest present-day black hole-to-stellar mass ratios, the largest stellar mass decrease is only $\approx0.28$ dex, while roughly half of these systems instead increase their stellar mass by up to $\approx0.32$ dex. Moreover, the median stellar mass evolution of the relic population is small ($\lesssim0.1$ dex for the low- and intermediate-ex-situ subsamples), despite substantially larger changes in galaxy size. These results indicate tidal stripping of stellar material is a negligible effect and does not drive  the elevated black hole-to-stellar mass ratios.} The observed offsets are therefore consistent with AGN feedback contributing to the rapid quenching of relic progenitors at $z\gtrsim 2$. This interpretation is also supported by observations of nearby relic galaxies, which host some of the most overmassive black holes known at fixed stellar mass \citep{van_den_Bosch_2012,Ferre-Mateu+15,Ferre-Mateu+17}. Within the two-phase formation scenario, such elevated ratios can arise when a galaxy assembles a substantial fraction of its black-hole mass during an early dissipative phase but subsequently avoids the stellar-mass growth that would otherwise reduce the black hole-to-stellar mass ratio.

However, the black-hole distribution reveals a more subtle trend. The different galaxy populations do not follow the simple expectation that relic galaxies should exhibit the largest black-hole excesses. Although SRGs are systematically offset above the median relation, ancient non-compact galaxies reach even larger black hole-to-stellar mass ratios. Compact galaxies with low DoR, in contrast, remain largely consistent with the median population. 

At first sight, the larger black hole-to-stellar mass ratios of non-compact SAGs relative to SRGs in COLIBRE may appear difficult to reconcile with the relic interpretation. However, the compactness threshold adopted in Section~\ref{sec:comp_crit} was derived empirically from changes in galaxy population properties and should not be regarded as a physically sharp boundary. Indeed, many SAGs classified as non-compact lie only slightly below the nominal threshold of \mbox{$\log_{10}(\Sigma_{1.5,\mathrm{crit}}/(\rm{M}_\odot \,\mathrm{kpc}^{-1.5})) = 9.75 \pm 0.10$}, with a median compactness of $\log_{10}(\Sigma_{1.5}/(\rm{M}_\odot \, \mathrm{kpc}^{-1.5})) \approx  9.5$ and a tail extending toward lower values. 
Moreover, the median ex-situ mass fraction of the non-compact SAGs is $f_\mathrm{ex-situ}= 0.019$ compared to $f_\mathrm{ex-situ}= 0.082$ for the SRGs. 
The stronger black-hole excess of non-compact SAGs therefore cannot be explained simply by a larger contribution from merger-driven stellar assembly. Instead, the distinction between SRGs and non-compact SAGs may be more gradual than implied by the adopted classification. Determining whether ancient non-compact galaxies represent a physically distinct population or the low-compactness tail of the SRG population will require a direct analysis of their formation, merger, and size-evolution histories. This will be the topic of the next paper of this series.  

More importantly, both SRGs and non-compact SAGs remain clearly separated from galaxies with low DoR. This indicates that black-hole excess is more strongly connected to early assembly and ancient stellar populations than to present-day compactness alone in COLIBRE. In this sense, ancientness appears to be a stronger predictor of elevated black-hole-to-stellar-mass ratios than structural compactness.

At the same time, the presence of overmassive black holes alone does not uniquely establish AGN feedback as the dominant quenching mechanism. The present analysis is restricted to galaxy properties at $z=0$ and does not directly follow black-hole histories or the temporal relation between AGN activity and the suppression of star formation. The observed offsets should therefore be interpreted as strong circumstantial evidence for an AGN-driven contribution to relic formation rather than definitive proof. 
\maxz{Nevertheless, an independent line of evidence is provided by the connection to the high-redshift massive quiescent galaxy (MQG) population studied by \citet{Chandro-Gomez+26a} and \citet{Chandro-Gomez+26b}. 89 per cent of these $z = 2$ MQGs already satisfy the compactness criterion \mbox{$\log_{10}(\Sigma_{1.5,\mathrm{crit}}/(\rm{M}_\odot\,\mathrm{kpc}^{-1.5})) = 9.75$} derived from the $z=0$ population, consistent with the interpretation that, when combined with a high DoR, this threshold selects galaxies that retain the structural imprint of the early formation phase. We find that 87 out of the 267 (32.6\%) relic candidates identified here are immediate descendants of the MQGs analysed in that work, which were shown to be predominantly quenched through AGN feedback. These 87 galaxies are primarily (65.5\%) satellite galaxies, have modest ex-situ fractions ($\lesssim 0.1$), and have characteristic stellar masses of $\approx 10^{10.5}\,\rm M_{\odot}$. They therefore represent particularly promising simulated counterparts of observed relic galaxies.  
The main progenitors at $z=2$ of the remaining 180 SRGs do not yet satisfy the MQG quenching criterion at that redshift. However, their median specific star formation rate of $\approx 2.39\times10^{-10}\,\mathrm{yr^{-1}}$ is already well below the main sequence at $z=2$, suggesting that a substantial fraction are in the process of quenching.
In addition, 55 of the 223 (24.7\%) non-compact SAGs descend from the MQGs identified by \citet{Chandro-Gomez+26a} at $z = 2$. 
Their stellar masses and ex-situ fractions are comparable to those of the relic descendants, while their median compactness ($\log_{10}(\Sigma_{1.5}/(M_{\odot}\,\rm kpc^{-1.5}))\approx9.5$) 
lies only slightly below the adopted compactness threshold. These galaxies also contain a substantially larger fraction of centrals (34.5\%) than the remaining non-compact SAG population (10.1\%).
Together, these results suggest that the high-$z$ MQG population in COLIBRE represents one important progenitor channel of present-day relic candidates, while also producing ancient galaxies that narrowly fail the adopted compactness criterion. This further supports the interpretation that the transition between non-compact SAGs and SRGs is continuous, rather than reflecting two fundamentally distinct galaxy populations. 
However, subsequent merger-driven growth naturally moves the majority of the MQG population at $z=2$ out of the compact relic regime by $z=0$, leaving only a minority that survive as present-day relic candidates in COLIBRE.} 



\subsection{Environmental dependence: central and satellite galaxies}

\begin{figure}
\centering
    \includegraphics[width=\columnwidth,
    trim={0.2cm 0cm 0cm 0cm}]{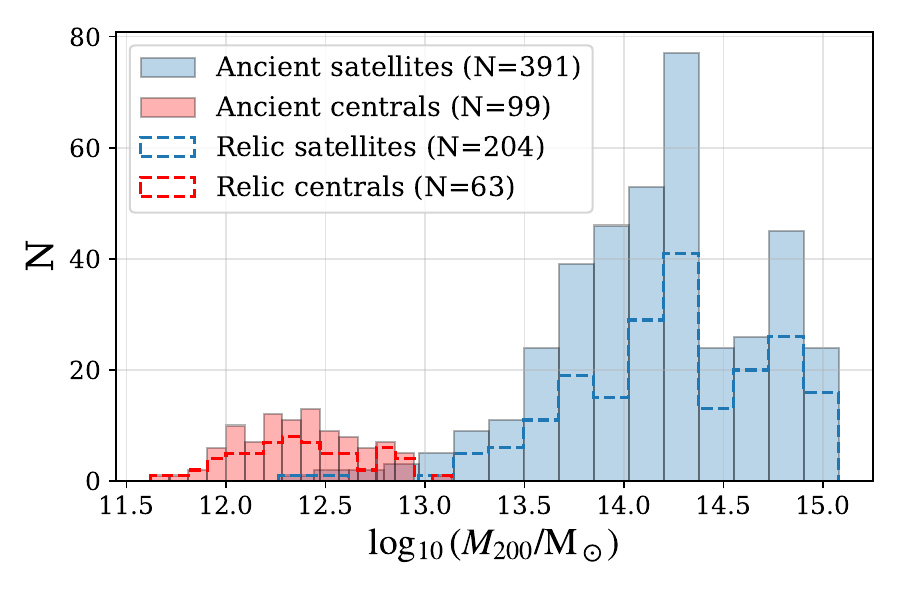}
    \caption{\maxz{Host halo mass ($M_{200}$) distribution of ancient (filled histograms) and relic (dashed histograms) galaxies. Satellite galaxies are shown in blue and centrals in red. Both ancient and relic satellites preferentially inhabit massive cluster-scale haloes, whereas centrals are found predominantly in lower-mass host haloes.}}
    \label{fig:hosthalomass} 
\end{figure}

The environments of SRGs provide an additional clue to their formation pathways. In particular, distinguishing between central and satellite systems allows us to assess whether relic-like evolution is primarily driven by internal processes or their environment could also be playing a role.

The majority of SRGs (204 out of 267) are satellites rather than central galaxies. This immediately suggests that environmental quenching mechanisms may play an important role in maintaining \maxz{and possibly creating} relic-like properties, especially at the massive end of the population. Satellite galaxies residing in dense environments are more likely to experience processes such as gas \maxz{and tidal} stripping, \maxz{which can make them compact and more ancient}, suppression of fresh gas accretion, and dynamical interactions that inhibit renewed star formation.

This interpretation is reinforced by the host halo masses shown in Figure~\ref{fig:hosthalomass}. Satellite SRGs preferentially inhabit substantially more massive host haloes than their central counterparts, with typical differences of approximately two orders of magnitude. The stellar mass range of these satellite SRGs is more extended and skewed to slightly lower values than the central SRGs (see red and blue stars in Figure~\ref{fig:BHmass}), showing that the tail of SRGs to stellar masses $\approx 10^{10}\,\rm M_{\odot}$ is driven by ancient and compact satellite galaxies. 
The halo mass distribution of Figure~\ref{fig:hosthalomass} for satellite SRGs shows an overrepresentation of galaxy clusters, in which  
environmental suppression of late star formation may help preserve compact stellar structures despite continued hierarchical assembly \maxz{of the host halo}. The extreme orbital velocities in cluster environments also help prevent galaxy mergers between satellite galaxies, preserving the compact nature of the SRGs. 

\begin{figure}
    \centering
    \includegraphics[width=\linewidth]{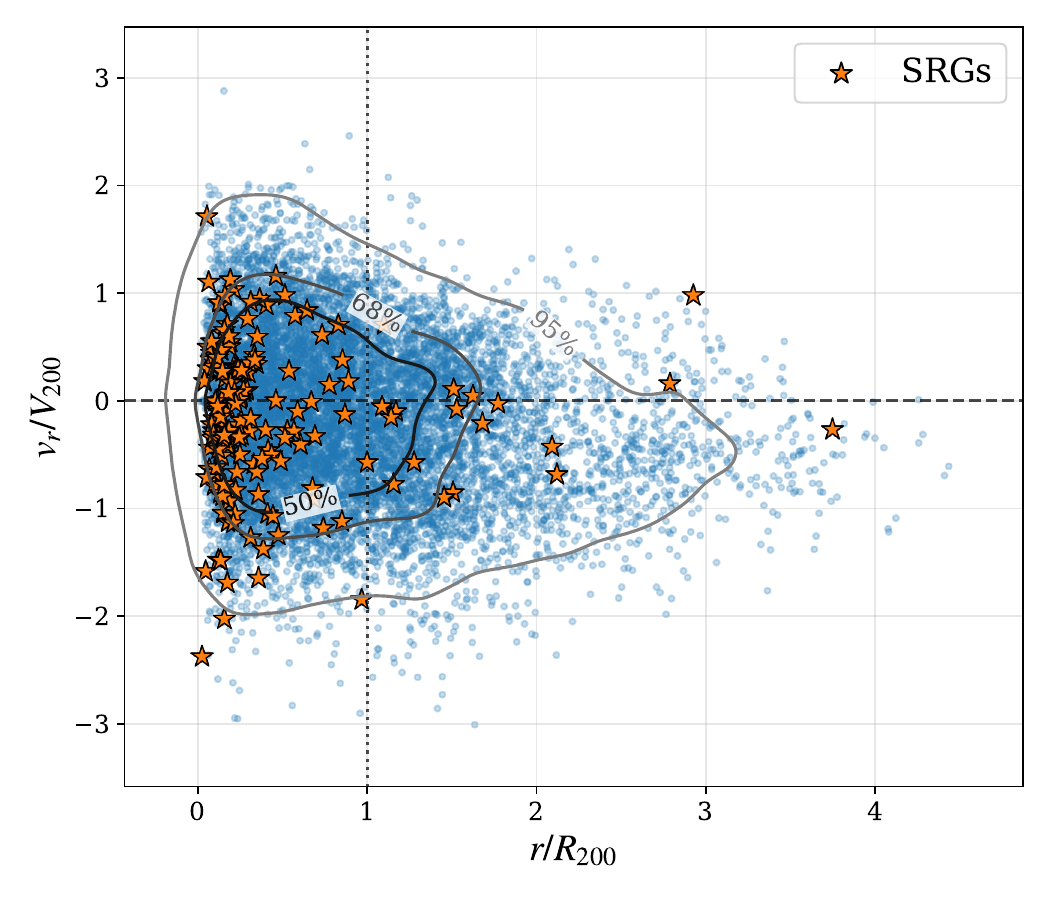}
    \caption{Projected relative velocities of galaxies with respect to the host velocity dispersion (virial velocity of the halo), plotted against the relative distance over $R_{200}$. All clusters with mass above $10^{14} \, \rm{M}_\odot$ are displayed, with their relics being highlighted with star symbols.}
    \label{fig:vrel-relicclusters}
\end{figure}

\maxz{Given the observational results from \INSPIRE\ \citep{Scognamiglio24}, which show that relic candidates preferentially inhabit the inner regions of galaxy clusters, we further explore the population of satellite SRGs to investigate whether they occupy similarly preferred regions of their host halo phase space.}
We restrict the analysis to host haloes with masses $M_{\rm host}>10^{14}\,\rm{M}_\odot$. This threshold serves two purposes. First, it focuses on the cluster regime, where environmental quenching processes are generally expected to be most efficient (e.g. \citealt{Peng2010, Peng2012, Wetzel2013, Cortese+21}). Second, it ensures a statistically meaningful sample of ancient and relic galaxies, whereas at lower host-halo masses their numbers become too small for a robust phase space analysis. As shown in Figure~\ref{fig:hosthalomass}, ancient and relic candidates above this mass scale are also almost exclusively satellites, making the distinction between internal and environmental quenching particularly relevant. In Figure~\ref{fig:vrel-relicclusters}, we further plot the relative velocities of the galaxies with respect to the virial velocity of their host clusters against the relative distance over the virial radius for all halos with $M_\text{host} > 10^{14} \, \rm{M}_\odot$, and highlight the SRGs among them. SRGs cluster around the centres of their host halos, $r/R_{200} < 0.2$, with velocities indicating virialisation, in qualitative agreement with observations. 
Such small cluster-centric distances are typically associated with satellites that have spent a significant fraction of their lifetime within the host halo (e.g. \citealt{Walker26}). The concentration of SRGs toward these inner regions therefore suggests that environmental effects are not merely acting at the present epoch, but may have influenced their evolution over extended periods. Early infall into massive clusters would naturally suppress fresh gas accretion, reduce the likelihood of rejuvenated star formation, and help preserve the compact and quenched nature of relic systems until the present day. This reinforces the conclusion drawn from the substantial host-halo mass difference between central and satellite relic candidates. Satellite relics, in particular, seem to not just reside in cluster environments,
but to preferentially populate the denser inner regions within those halos. 

An alternative explanation is that these satellite galaxies are accreted early on and suffer from significant tidal stripping, leaving only the compact core of the initial galaxy. The outcome would be a compact, ancient galaxy, even though in this case the galaxy would not necessarily have been compact, nor globally ancient before falling into the cluster. We tested this scenario by tracing back the compactness of SRGs over $\approx 13 \,\mathrm{Gyr}$. 
\maxz{Only about $3\%$ of the SRGs exhibit a clear increase in compactness over time, which would be a smoking gun for significant tidal stripping. Instead, the compactness of the vast majority of SRGs remains remarkably stable over the last $\approx 10 \,\mathrm{Gyr}$: between $z=1.5$ and $z=0$ the median compactness decreases by only $\approx0.3-0.4$ dex, while the median half-mass radius increases by only $\approx0.2-0.35$ dex, depending on ex-situ fraction. Together with the nearly unchanged stellar masses, these trends indicate that most satellites were already compact before entering their present-day environments and that tidal stripping plays at most a minor role in producing the present-day SRG population.}
%
A detailed analysis of their evolutionary histories will be presented in a follow-up paper.

In summary, the environmental trends indicate that SRGs do not share a single environmental history. Intermediate-mass SRGs with low ex-situ fractions are broadly consistent with the classic picture of untouched relics formed through rapid early assembly and minimal subsequent growth. However, the predominance of satellites among SRGs, particularly in the inner regions of massive clusters, suggests that environmental processes may subsequently help maintain quiescence and suppress rejuvenation. The stability of their compactness histories nevertheless indicates that cluster environments do not generally produce these systems through tidal stripping; for most SRGs, compactness was established at early times and remained largely unchanged. Environment therefore appears to contribute primarily to the preservation, rather than the creation, of compact quenched systems.


\section{Discussion}\label{sec:discussion}

\subsection{The stellar mass--size plane as a fossil record of galaxy assembly}
One of the central results of this work is that the present-day structure of galaxies, and in particular their compactness, retains measurable information about their evolutionary histories. 
Across the COLIBRE galaxy population, independent tracers of stellar populations, star formation histories, and galaxy dynamics exhibit coherent changes across a similar compactness range. 
The stellar mass--size plane therefore contains an evolutionary dimension in addition to galaxy mass and size (Figure~\ref{fig:fourpanels_masssize}). Rather than tracing morphology alone, compactness is linked to both the timing and duration of stellar formation and the subsequent assembly of galaxies.
This behaviour emerges consistently from six complementary diagnostics. Luminosity-weighted stellar age, stellar [Mg/Fe], specific star formation rate, stellar metallicity, stellar velocity dispersion, and ex-situ stellar mass fraction all identify a characteristic transition at
$\log_{10}(\Sigma_{\rm 1.5,crit}/(\rm M_{\odot}\,\mathrm{kpc}^{-1.5}))=9.75\pm0.10$ 
(Figure~\ref{fig:fourpanels_masssize_1d}). Their agreement provides a physically motivated compactness criterion, rather than one based solely on the empirical structural cuts commonly adopted in observational studies (e.g. \citealt{Barro13}).
The consistency among these independent tracers suggests that the derived threshold marks a physically meaningful change in the average evolutionary properties of galaxies. It should nevertheless be interpreted as a characteristic transition scale rather than as a sharp physical boundary between compact and non-compact systems.
An important caveat, however, is that our derived compactness threshold is based on the intrinsic three-dimensional stellar half-mass radius. It therefore cannot be translated directly into an observational selection based on projected half-light radii, which additionally depend on stellar-population gradients, mass-to-light ratio variations, projection effects, and the adopted photometric modelling. A direct observational calibration of the threshold will require constructing projected mock observables from the simulation. 

We therefore propose {\it morpholography} as a framework in which present-day galaxy structure is interpreted as an observable fossil record of assembly history. In this picture, the stellar mass--size plane no longer represents only morphology, \maxz{but also provides access to the evolutionary information encoded in present-day stellar populations, thereby offering indirect insight into the physical processes that shaped galaxies throughout cosmic time.}

\subsection{Compactness and ancientness do not individually imply relicness}
A second important result is that neither structural compactness nor an ancient stellar population alone uniquely identifies a relic galaxy. In our terminology, simulated compact galaxies (SCGs) are selected solely through the compactness criterion, whereas simulated ancient galaxies (SAGs) are selected through \maxz{a Degree of Relicness of} $\text{DoR}>0.6$, independently of their present-day structure. Simulated relic galaxies (SRGs) are defined as the overlap between these two populations.

Applying the observationally motivated DoR to the full simulated galaxy population shows that SCGs span a broad range of DoR values and are therefore not uniformly ancient. This remains true despite the tendency of compact galaxies to host older stellar populations and exhibit enhanced [Mg/Fe], suppressed star formation, and higher stellar velocity dispersions. Moreover, their ex-situ stellar mass fractions span almost the full range (Figure~\ref{fig:fourpanels_masssize_1d}f), revealing a diversity of merger-driven assembly histories. Compactness is therefore closely associated with early stellar formation, but does not uniquely determine either ancientness or subsequent assembly history. 
Conversely, a substantial population of SAGs lies below the compactness threshold. These galaxies formed their stellar populations early and over short timescales, \maxz{but do not satisfy our present-day compactness criterion.} SRGs therefore occupy only the intersection between structural compactness and high DoR, rather than being synonymous with either SCGs or SAGs individually.

The ex-situ mass fraction introduces a further distinction within this overlap. Compact high-DoR systems with low ex-situ fractions most closely resemble classical untouched relics, whereas SRGs with substantial ex-situ contributions have experienced more complex merger-driven assembly despite retaining compact present-day structures. Ex-situ fraction is not part of our formal SRG selection, but provides an important diagnostic of the diversity of assembly histories within the relic-candidate population.

These results help explain why selections based on compactness or ancient stellar populations alone recover mixed evolutionary populations, and reinforce the need to combine structural and evolutionary information when identifying relic candidates.

\subsection{Relic galaxies follow multiple evolutionary pathways}

Perhaps the most surprising outcome of this work is that relic galaxies themselves do not appear to arise through a single evolutionary pathway.

The classical (observation-based) picture describes relic galaxies as systems that formed rapidly at high redshift and subsequently experienced little additional growth. While COLIBRE indeed predicts such galaxies, it also reveals a second population of compact systems with similarly ancient stellar populations but substantially larger ex-situ stellar mass fractions. These galaxies assembled a significant fraction of their stellar mass through mergers that occurred at $z>2$. 

Present-day relic candidates in COLIBRE appear to comprise at least two distinct evolutionary channels: classical relics dominated by early in-situ formation, and early merger-built ancient systems assembled from stars that themselves formed predominantly at high redshift.

The importance of these channels depends on stellar mass. "Classical" relics are most common around stellar masses of $10^{10}-10^{10.5}\,\rm M_{\odot}$, whereas more massive compact galaxies increasingly exhibit large ex-situ stellar mass fractions (Figure~\ref{fig:relicsample}), suggesting that merger-driven assembly becomes progressively more important at the highest masses.
The characteristic stellar masses of the SRGs identified in COLIBRE are somewhat lower than those of most observationally confirmed relic galaxies, which are commonly concentrated around $\approx10^{11}\,\rm{M}_\odot$ \citep[see, e.g., fig.~3 of][]{Maksymowicz-Maciata24}. 
However, Such searches often focus on pre-selected samples of massive and/or ultra-compact galaxies \mbox{($\log_{10}(M_\star/\rm M_\odot)\gtrsim10.5$)}\footnote{Although
\citet{Grebor23} also identify compact, ancient systems at
$\log_{10}(M_\star/\rm M_\odot)\approx10.0$.}.  
By contrast, our analysis applies the compactness and DoR criteria to the full simulated galaxy population above $10^9\,\rm{M}_\odot$, allowing lower-mass relic-like systems to emerge without an a priori stellar-mass selection. 
A direct quantitative comparison is further complicated by differences in the structural quantities used. The compactness criterion adopted here is based on intrinsic three-dimensional stellar half-mass radii, whereas observational studies generally rely on projected half-light radii. The characteristic mass distribution and compactness threshold derived here should therefore not be interpreted as directly equivalent to those obtained from observational selections. A robust comparison with observational relic samples will require mock-observational measurements of the simulated galaxies.

The relic population also displays distinctive present-day properties. Relic candidates preferentially host black holes that are approximately $0.25-2.5$~dex more massive than typical galaxies of the same stellar mass (Figure~\ref{fig:BHmass}), are predominantly satellites residing within massive cluster environments (Figure~\ref{fig:hosthalomass}), and approximately one third are direct descendants of the massive quiescent galaxy population already in place by $z=2$, analysed in \citet{Chandro-Gomez+26a,Chandro-Gomez+26b} in COLIBRE. \maxz{While the majority of high-redshift MQGs subsequently evolve out of the compact relic regime through merger-driven growth, these descendants represent the subset that preserves both their early quenching history and compact structure until the present day. Together, these results place present-day relic candidates within a continuous evolutionary framework linking the earliest quenched galaxies to the surviving compact population at $z=0$.}

\section{Conclusions}
\label{sec:conclusions}

In this work, we used the COLIBRE cosmological hydrodynamical simulations, in particular the L200m6 run, to investigate whether the present-day structure of galaxies retains information about their stellar formation and assembly histories, and to establish a physically motivated framework for identifying relic galaxies. 
The principal conclusions of this work are:

\begin{itemize}
    \item {\bf The stellar mass--size plane contains an evolutionary dimension in addition to galaxy mass and size}. \maxz{Independent tracers of stellar populations (age, metallicity, [Mg/Fe]), star formation histories (specific star formation rate), galaxy dynamics (stellar velocity dispersion), and ex-situ assembly all converge on the same characteristic compactness scale, demonstrating that present-day galaxy structure acts as an observable fossil record of galaxy formation and assembly (see Figure \ref{fig:fourpanels_masssize}).} We refer to this mapping between structure and evolutionary history as {\it morpholography}, providing a new framework for linking nearby galaxies to the physical processes that shaped them in the early Universe.
    \item {\bf A characteristic compactness threshold emerges from multiple galaxy evolution diagnostics}. The six complementary tracers listed above independently identify a common characteristic compactness of $\log_{10}(\Sigma_{\rm 1.5,crit}/(\rm M_{\odot}\,\rm kpc^{-1.5}))=9.75\pm 0.10$ (see Figure \ref{fig:fourpanels_masssize_1d}). The convergence of these otherwise distinct diagnostics suggests that this scale marks a physically meaningful change in the average evolutionary properties of galaxies, providing an alternative to the empirical compactness cuts commonly adopted in observational studies. The threshold should nevertheless be interpreted as a characteristic transition scale rather than as a sharp physical boundary.
    \item {\bf Neither compactness nor ancientness alone is sufficient for relicness}. Simulated compact galaxies (SCGs), selected solely through their present-day structure, \maxz{span a broad range of Degree of Relicness (DoR; Eq. \ref{DoR}) values, where the DoR provides a quantitative measure of the star formation history, and ex-situ stellar mass fractions (see Figure \ref{fig:3panel_DoRvsquantity}, \ref{fig:panel-DoRvsexsitu}).} Conversely, a substantial population \maxz{(45.5\%)} of simulated ancient galaxies (SAGs), selected through $\text{DoR}>0.6$, lies below the compactness threshold. High DoR therefore identifies early stellar formation, but does not by itself imply the preservation of a compact structure or the avoidance of subsequent merger-driven growth. Simulated relic galaxies (SRGs) occupy the overlap between the SCG and SAG populations \maxz{(i.e. satisfying both $\text{DoR}>0.6$ and $\log_{10}(\Sigma_{\rm 1.5,crit}/(\rm M_{\odot}\,\rm kpc^{-1.5}))>9.75$}), demonstrating that structural and evolutionary information must be combined to identify relic candidates.
    \item {\bf Ancient stellar populations and relics arise through multiple evolutionary pathways.} Compact high-DoR galaxies with low ex-situ fractions most closely follow the classical relic scenario of rapid early in-situ formation followed by minimal subsequent growth. Other ancient systems contain substantial ex-situ stellar components, showing that stars can form early in external progenitors and be accreted later without erasing their ancient stellar-population signatures. Relic candidates are preferentially found at stellar masses of $\approx 10^{10}-10^{11}\,\rm M_{\odot}$, and host black holes approximately $0.25-2.5$~dex more massive than for typical galaxies of the same stellar mass (see Figure \ref{fig:BHmass}). Approximately $76\, \%$ of SRGs are satellites, and cluster SRGs preferentially occupy the inner regions of their host haloes (see Figure \ref{fig:vrel-relicclusters}). However, only $\approx3\, \%$  show a clear increase in compactness consistent with strong tidal stripping, indicating that environment primarily helps preserve quiescence rather than creating compact remnants. Finally, 87 of the 267 SRGs, approximately one third of the population, are direct descendants of the $z=2$ massive quiescent galaxies identified in COLIBRE, supporting an evolutionary connection between early AGN-quenched systems and present-day relic candidates. 
\end{itemize}

The framework developed here provides several observationally testable predictions. The \INSPIRE\ and \EINSPIRE\ surveys \citep{Spiniello+24,Mills25}, together with large integral-field spectroscopic surveys and future facilities, provide suitable datasets with which to test the predicted connections between compactness, stellar populations, black-hole mass, and environment. A direct observational calibration of the compactness threshold will require projected mock measurements based on half-light radii, rather than the intrinsic three-dimensional half-mass radii used in this work. More broadly, {\it morpholography} provides a framework for interpreting present-day galaxy structure not merely as morphology, but as a measurable fossil record of galaxy formation and assembly.

\section*{Acknowledgements}

This work used the DiRAC@Durham facility managed by the Institute for Computational Cosmology on behalf of the STFC DiRAC HPC Facility (www.dirac.ac.uk). The equipment was funded by BEIS capital funding via STFC capital grants ST/P002293/1, ST/R002371/1 and ST/S002502/1, Durham University and STFC operations grant ST/R000832/1. DiRAC is part of the National e-Infrastructure. The work also used the Hyades supercomputer at ICRAR/UWA and we thank Mark Boulton for his continued efforts in managing the machine.\\
\maxz{FH acknowledges funding from the Netherlands Organization for Scientific Research (NWO) through research programme Athena 184.034.002. EC acknowledges support from STFC consolidated grant ST/X001075/1. SP acknowledges support by the Austrian Science Fund (FWF) through grant-DOI: 10.55776/V982. MS acknowledges the support from the Swiss National Science Foundation under grant no. P5R5-2\_239162.}
\section*{Data Availability}

The data supporting the plots within this article are available on reasonable request to the corresponding author. The COLIBRE  simulation data will eventually be made publicly available, although we note that the data volume (several petabytes) may prevent us from simply placing the raw data on a server. In the meantime, people interested in using the simulations are encouraged to contact the corresponding author. 



\bibliographystyle{mnras}
\bibliography{bibliography} %



\FloatBarrier
\clearpage
\appendix

\section{Final assembly time}
\label{app:tfin}
\begin{figure}
    \centering
    \includegraphics[width=\linewidth]{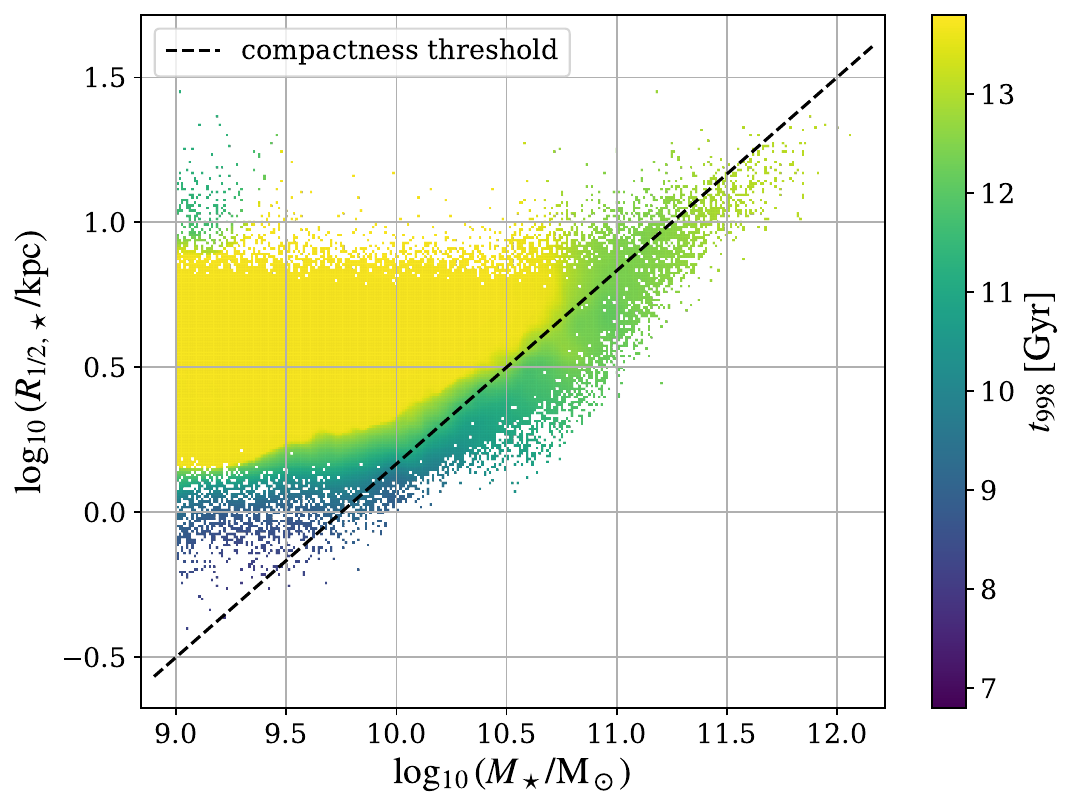}
    \caption{Stellar mass--size plane colour-coded by the cosmic time at which 99.8\% of the final stellar mass is in place ($t_{99.8}$). The dashed line marks the compactness threshold derived in Section~\ref{sec:comp_crit}.}
    \label{fig:Mass-size_t998}
\end{figure}

\chiara{The DoR includes the term $t_{\rm fin}$, referred to as the
``final assembly time'' in the original observational definition. Operationally, it is derived from the cumulative star-formation
history: it records when the stars formed, not when ex-situ stars
were accreted. In the \INSPIRE\ definition, $t_{\rm fin}$ is the
cosmic time when cumulative formed stellar mass reaches 99.8\%
of its final value.} 
When applied to the COLIBRE galaxies, however, the 99.8\% definition becomes sensitive to the finite sampling of the stellar population by star particles, particularly at the low-mass end. In these systems, a very small number of late-forming star particles can account for more than 0.2\% of the final stellar mass and therefore shift $t_{\mathrm{fin}}$ to cosmic times close to the present day. This effect is visible in Figure~\ref{fig:Mass-size_t998}, where a broad region of the low-mass galaxy population is assigned $t_{998}$ values close to the age of the Universe. The limited dynamic range in this regime primarily reflects particle sampling rather than meaningful differences in the duration of stellar mass formation.

\begin{figure}
    \centering
    \includegraphics[width=\linewidth]{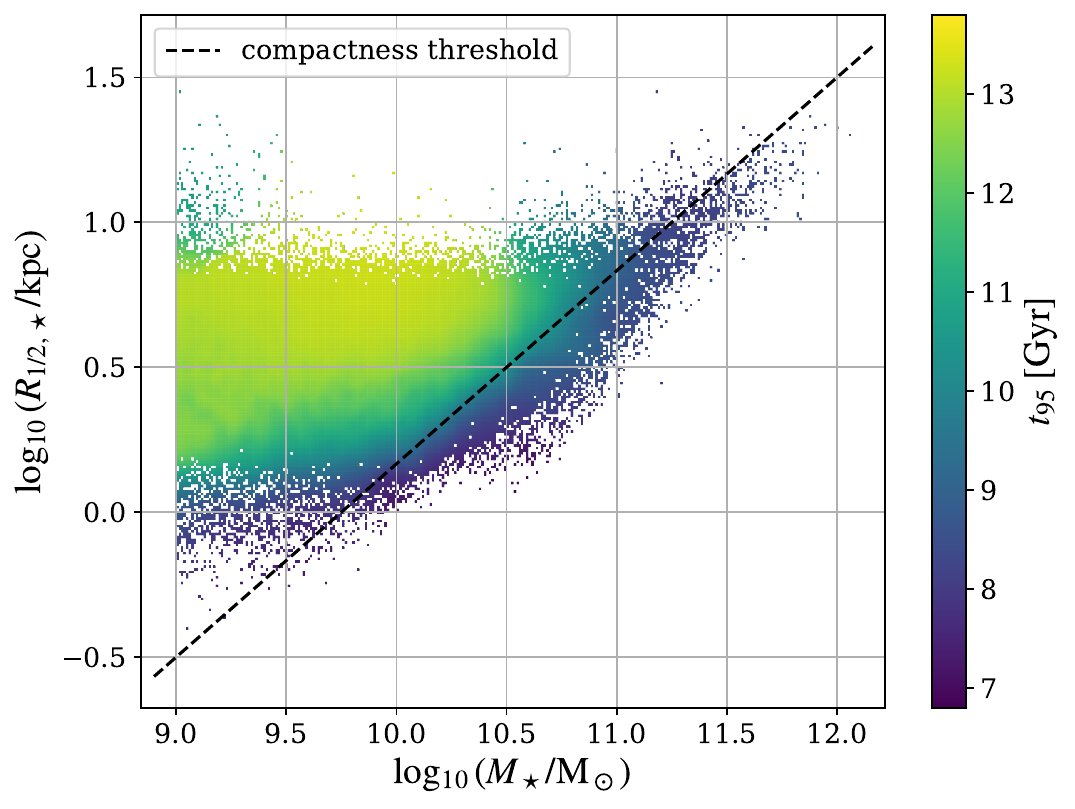}
    \caption{Stellar mass--size plane colour-coded by the cosmic time at which 95\% of the final stellar mass is in place ($t_{95}$). The dashed line marks the compactness threshold derived in Section~\ref{sec:comp_crit}.}
    \label{fig:Mass-size_t95}
\end{figure}

We therefore define $t_{\mathrm{fin}}$ as the epoch at which 95\% of the final stellar mass is in place. This larger tolerance reduces the influence of individual late-forming star particles and produces a smoother variation across the stellar mass--size plane, as shown in Figure~\ref{fig:Mass-size_t95}. Galaxies with genuinely extended formation histories still reach late values of $t_{95}$, but the widespread artificial accumulation near the present epoch is substantially reduced. Importantly, the broad dependence of $t_{\mathrm{fin}}$ on stellar mass and structure remains qualitatively unchanged.

This modification affects only the final assembly-time term of the DoR; the fraction of stellar mass formed by $z=2$, $f_{M^*_{z=2}}$, and the formation time $t_{75}$ remain unchanged. The adoption of $t_{95}$ should therefore be regarded as a numerical adaptation of the observational definition to the mass resolution of COLIBRE, rather than as a change in the underlying physical interpretation of the DoR.

\section{Relic morphology}
\label{app:morphology}
\begin{figure*}
    \centering    \includegraphics[width=0.75\linewidth]{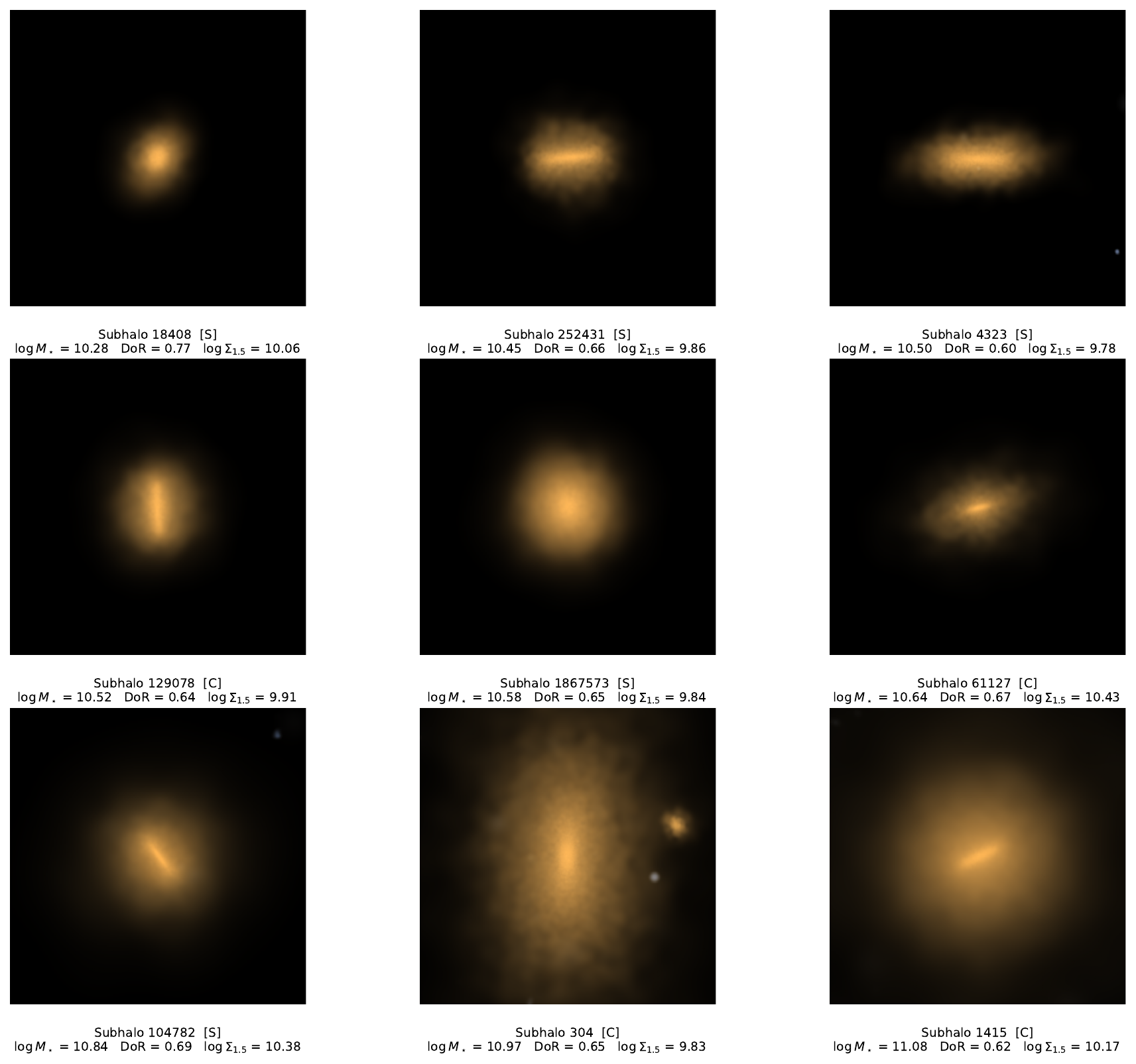}
    \caption{Examples of SRGs with typical isolated morphologies \maxz{ shown at random orientation}. The sample includes both central [C] and satellite [S] galaxies and is ordered by increasing stellar mass.}
    \label{fig:isolated_relics}
\end{figure*}

\begin{figure*}
    \centering
    \includegraphics[width=0.75\linewidth]{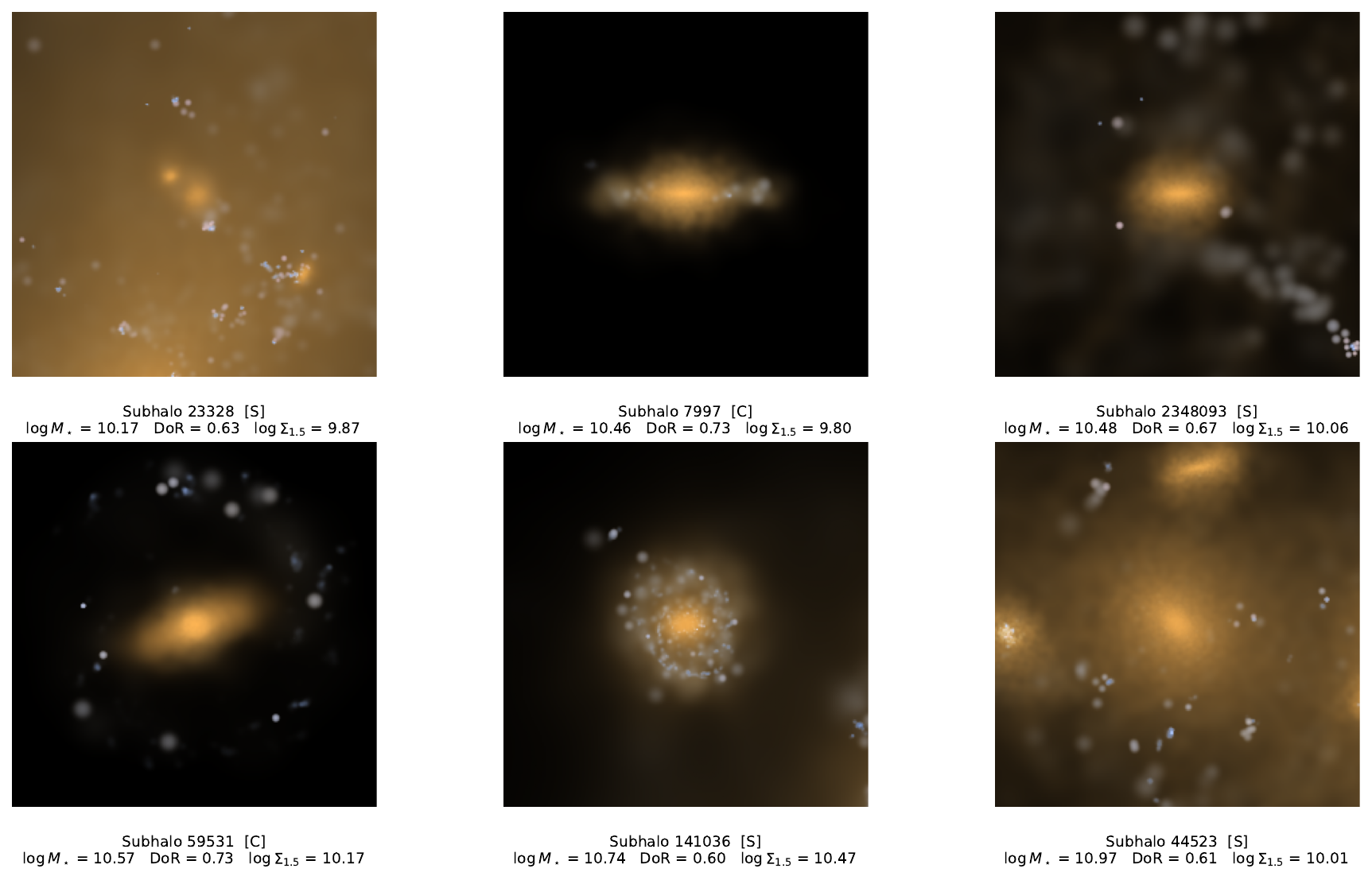}
    \caption{Examples of SRGs exhibiting nearby companions or surrounding structures that may indicate future mergers or ongoing matter infall. \maxz{A random orientation is used for visualisation.} The sample includes both central [C] and satellite [S] galaxies and is ordered by increasing stellar mass.}
    \label{fig:interacting_relics}
\end{figure*}

Figures~~\ref{fig:isolated_relics} and ~\ref{fig:interacting_relics} show a representative sample of 
simulated relic galaxies (SRGs) identified in Section~\ref{sec:compact-relic-relation}. \maxz{The images were generated from the COLIBRE simulation using the PARTRIDGE synthetic-imaging package \citep[in preparation;][]{Husko_inprep}, following the synthetic-image methodology described by \citet{Schaye+26}. We use the $\texttt{SDSS\_original}$ imaging mode to produce dust-free synthetic rest-frame SDSS $gri$ colour-composite images at $z=0$, in which the $g$, $r$, and $i$ filter fluxes are mapped directly to the blue, green, and red RGB channels, respectively. The images cover a physical field of view of $50\,\mathrm{kpc}\times50\,\mathrm{kpc}$ centred on each galaxy. 
Random projections are shown for each system.} 
The first figure includes isolated relics that morphologically resemble observed nearby relics (e.g., NGC~1277 from \citealt{Trujillo14} or J0842 from \citealt{Tortora25}) while the second focusses on systems that show evidence for future mergers or signs of gas infall or other interactions, and hence might not be relics in the future, although being compact and ancient today.

\bsp	
\label{lastpage}
\end{document}